\documentclass[twocolumn,aps,prb,showpacs,amsmath,superscriptaddress,longbibliography,notitlepage]{revtex4-1}
\usepackage{amssymb}
\usepackage{mathrsfs}
\usepackage{graphicx}
\usepackage{float}
\usepackage[caption=false]{subfig}
\usepackage[pdftex,colorlinks=red]{hyperref}
\usepackage{makecell}

\usepackage{epstopdf}
\usepackage[flushleft]{threeparttable}
\usepackage{booktabs}

\usepackage[normalem]{ulem}
\usepackage{verbatim}
\usepackage{xcolor}
\usepackage{bm}

\def\red{\textcolor{red}}

\bibpunct{[}{]}{,}{n}{}{}

\begin{document}

\title{Topological properties of non-Hermitian Creutz Ladders}

\author{Hui-Qiang Liang}
\affiliation{Guangdong Provincial Key Laboratory of Quantum Metrology and Sensing $\&$ School of Physics and Astronomy, Sun Yat-Sen University (Zhuhai Campus), Zhuhai 519082, China}
\author{Linhu Li}  \email{lilh56@mail.sysu.edu.cn}
\affiliation{Guangdong Provincial Key Laboratory of Quantum Metrology and Sensing $\&$ School of Physics and Astronomy, Sun Yat-Sen University (Zhuhai Campus), Zhuhai 519082, China}

\begin{abstract}
In this work we study topological properties of the one-dimensional Creutz ladder model with different non-Hermitian asymmetric hoppings and on-site imaginary potentials, and obtain phase diagrams regarding the presence and absence of an energy gap and in-gap edge modes. The non-Hermitian skin effect (NHSE), which is known to break the bulk-boundary correspondence (BBC), emerges in the system only when the non-Hermiticity induces certain unbalanced non-reciprocity along the ladder.  The topological properties of the model are found to be more sophisticated than that of its Hermitian counterpart, whether with or without the NHSE. In one scenario without the NHSE, the topological winding is found to exist in a two-dimensional plane embedded in a four-dimensional space of the complex Hamiltonian vector. The NHSE itself also possesses some unusual behaviors in this system, including a high spectral winding without the presence of long-range hoppings, and a competition between two types of the NHSE, with the same and opposite inverse localization lengths for the two bands, respectively. Furthermore, it is found that the NHSE in this model does not always break the conventional BBC,  which is also associated with whether the band gap closes at exceptional points under the periodic boundary condition.
\end{abstract}

\date{\today}

\maketitle

\section{introduction}
Topological phases of matter are fascinating as they host robust in-gap boundary modes protected by the topological properties of bulk bands under the periodic boundary condition (PBC), known as the bulk-boundary correspondence (BBC)\cite{hasan2010colloquium,qi2011topological,bernevig2013topological}.
In recent years, the scope of topological phases has been extended to non-Hermitian Hamiltonians, which are physically relevant as effective Hamiltonians of, for instance, open systems or wave systems with gain and loss \cite{NHbook,
Bender1998nonH,bender2007making,
Rotter2009non,
yoshida2018non,yamamoto2019theory,
ruter2010observation,longhi2018parity,ozawa2019topological}.
In non-Hermitian systems, the conventional topological BBC is modified by the non-Hermitian skin effect (NHSE), where all eigenmodes accumulate at the boundaries due to non-Hermitian pumping \cite{Lee2016nonH_Creutz,Martinez2018nonH_Creutz,xiong2018does,kunst2018biorthogonal,yao2018edge,yokomizo2019non,Lee2019anatomy}.
The topological modes under the open boundary condition (OBC) are now topologically characterized by a non-Bloch Hamiltonian in the so-called generalized Brillouin zone, with the crystal momentum replaced by a complex value \cite{yao2018edge,yokomizo2019non,Lee2019anatomy,li2019geometric,lee2020unraveling,yang2020non}.
On the other hand, the NHSE itself also represents a unique topological phenomenon in non-Hermitian systems, originated in a nontrivial spectral winding of the complex spectrum \cite{zhang2019correspondence,okuma2020topological,wang2021generating,li2021quantized}.

The Creutz ladder describes a one-dimensional (1D) two-leg ladder lattice with crosse-linked hoppings and a perpendicular magnetic flux \cite{Creutz1999}.
Representing one of the minimal 1D topological insulating systems, the Creutz ladder and its extensions have been employed to investigate various topological properties and phenomena \onlinecite{bermudez2009topology,Sticlet2014Creutz,Hugel2014Creutz,li2015hidden,Li2015Creutz,Lim2017Creutz,Junemann2017coldatom,kang2020creutz,zhou2020Creutz},
such as topological quench dynamics \cite{bermudez2009topology}, effects of superconductivity pairing on fractionally charged midgap states \cite{Sticlet2014Creutz}, 
cold atom realization of spin-orbit coupling and quantum Hall insulators \cite{Hugel2014Creutz},
and competitions between topological features and interaction effects in quantum many-body systems \cite{Junemann2017coldatom}, among many others. 
Non-Hermitian extensions of the Creutz ladder have also been introduced in investigating anomalous edge localization of eigenmodes (i.e. the NHSE) \cite{Lee2016nonH_Creutz,Martinez2018nonH_Creutz} and
non-Hermitian topological characterizations and BBC \cite{jin2019nonH_Creutz,wu2019nonH_Creutz,lee2020nonH_Creutz,lee2020nonH_Creutz,zhou2021dual},
and a potential application of controlling the NHSE via lattice shaking \cite{li2020topological}. 

In this paper we provide a comprehensive investigation of the effects of two types of non-Hermticity acting on the Creutz ladder model, namely on-site gain and loss represented by imaginary potentials, and hoppings with asymmetric amplitudes toward opposite directions.
We find that the NHSE is absent in some of the cases with asymmetric hoppings, as the induced non-reciprocities for different parts of the system are balanced and the system becomes reciprocal. 
Nevertheless, the topological characterization and transition are still distinct from that of Hermitian systems.
In other cases with the presence of the NHSE, we unveil that the conventional BBC is not always violated in the presence of the NHSE and a nontrivial spectral winding.
Furthermore, in our model, 
the NHSE induced by the imaginary on-site potential can be classified into two types, 
with the same and opposite skin localizations for the two bands, respectively, and the overall NHSE results from their competition.
These scenarios with different non-Hermitian parameters are carefully examined in this work, and their topological transitions and phase diagrams are obtained analytically and/or numerically.

The rest of the paper is organized as follows.
In Sec. \ref{II}, we set the stage of this paper by introducing the topological properties of the Creutz ladder model and the non-Hermitian parameters we considered.
We then discuss the topological properties of our model in the absence and presence of the NHSE, in Sec. \ref{III} and Sec. \ref{IV} respectively. Finally, our results are briefly summarized in Sec. \ref{V}.

\section{Model Hamiltonian}\label{II}
The fermionic Creutz ladder is described by the Hamiltonian \cite{Creutz1999}
\begin{eqnarray}
H&=&\sum_x^L \left[K(e^{i\theta}\hat{a}_{x}^\dagger\hat{a}_{x+1}+e^{-i\theta}\hat{b}_{x}^\dagger\hat{b}_{x+1})\right.\nonumber\\
&&\left.+Kr(\hat{a}_{x}^\dagger\hat{b}_{x+1}+\hat{b}_{x}^\dagger\hat{a}_{x+1})+M\hat{a}_{x}^\dagger\hat{b}_{x}\right]+\text{h.c.},\label{eq:CL_Hermitian}
\end{eqnarray}
where $K$, $r$, $M$, and $\theta$ are parameters controlling the amplitudes and phases of the hoppings, $L$ is the number of unit cells, and we will set $K=1$ as the energy unit in the rest of this paper. 
The Bloch Hamiltonian reads $\hat{H}(k)=\hat{\Psi}_k^\dagger h(k)\hat{\Psi}_k$ with $\hat{\Psi}_k^\dagger=(\hat{a}^\dagger_k,\hat{b}^\dagger_k)$ and
\begin{eqnarray}
h(k)&=&h_0\tau_0+h_z\tau_z+h_x\tau_x,\nonumber\\
h_0&=&2\cos k\cos \theta,~~
h_z=-2\sin k\sin\theta,\nonumber\\
h_x&=&M+2r\cos k,
\end{eqnarray}
with $k$ the lattice momentum, $\sigma_0$ and $\sigma_{x,y,z}$ the identity matrix and the Pauli's matrices acting on the sublattice space.
Here and in the following discussion, the $k$-dependence in the coefficients $h_{0,x,y,z}$ is omitted in their notations for simplicity.

\begin{figure}
\includegraphics[width=1\linewidth]{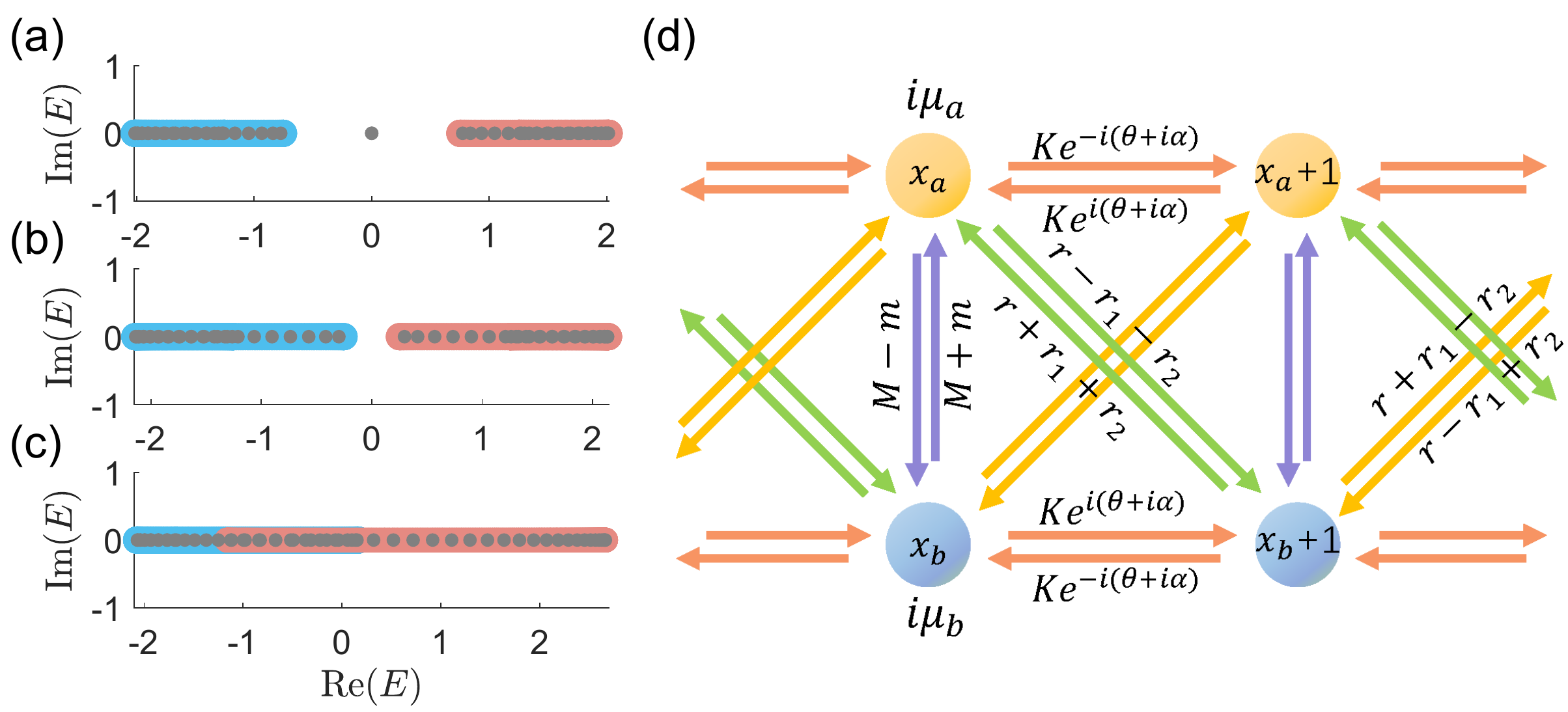}
\caption{
The Hermitian Creutz ladder model of Eq. \eqref{eq:CL_Hermitian} has three different topological phase, (a) a topologically non-trivial phase with a pair of degenerate in-gap edge modes, (b) a topologically trivial phase with a band gap, and (c) a gapless phase where the two bands overlaps. Blue and red colors indicate the two PBC bands, and gray dots are the OBC eigenenergies.
The parameters are (a) $M=0.25$, $r = 0.5$, $\theta=\pi/2$, (b) $M=0.75$, $r = 0.25$, $\theta=\pi/2$, (c) $M=0.75$, $r = 0.25$, $\theta=\pi/4$, with $L=30$ unit cells.
A sketch of the Creutz ladder model with all non-Hermitian parameters considered in this paper is shown in (d). 
}
\label{fig:CL}
\end{figure}

At $\theta=\pm\pi/2$, $h(k)$ satisfies a chiral symmetry $\sigma_yh(k)\sigma_y=h(k)$, and is analogous to the Su-Schrieffer-Heeger (SSH) \cite{SSH} model after rotating the Pauli's matrices 
\begin{eqnarray}
\sigma_y\rightarrow\sigma_z\rightarrow-\sigma_y.\label{eq:rotation}
\end{eqnarray}
Yet it belongs to the AIII class as the original Hamiltonian does not satisfy the time-reversal symmetry either for spinful or spinless systems.
The model is thus topologically described by the winding of $(h_x,h_z)$ throughout the Brillouin zone (BZ),
and supports topologically distinguished phases with and without a pair of degenerate topological edge modes in the band gap [examples are shown Fig. \ref{fig:CL}(a) and (b)].
Another parameter regime with $\theta=0,\pi$ yields a non-topological system as the Pauli matrix $\sigma_z$ vanishes, and may only be gapped when $|M|>|2r|$.
In more general cases, the non-trivial winding of $(h_x,h_z)$ survives, 
but a nonzero $h_0(k)$ induces a $k$-dependent shifting of the spectrum, 
which may lead to the overlapping of the two bands and the absence of a band gap in certain parameter regimes, as shown Fig. \ref{fig:CL}(c).

Compared with the SSH model, the Creutz ladder has more tunable parameters, and the existence of a nonzero $k$-dependent $h_0$ term breaks the symmetry of the spectrum between $E$ and $-E$. 
Thus 
richer  phenomena may emerge when non-Hermiticity is introduced to different parameters of the Hamiltonian, as discussed in the following sections.
Specifically, we shall focus on two types of non-Hermiticity widely investigated in literatures, namely extra imaginary on-site potentials given by
\begin{eqnarray}
H_\mu=i\mu_a\hat{a}_x^\dagger\hat{a}_x+i\mu_b\hat{b}_x^\dagger\hat{b}_x,
\end{eqnarray}
and asymmetric amplitudes of the hoppings without changing their phase factor ($\theta$ for the intra-sublattice hoppings and $0$ for the rest), given by replacements of the parameters
\begin{eqnarray}
r\rightarrow r\pm r_1\pm r_2,~M\rightarrow M\pm m, \theta\rightarrow\theta+i\alpha
\end{eqnarray}
with $r_{1,2}$, $m$, and $\alpha$ taking real values, and $\pm$ corresponding to the hoppings toward opposite direction.
Note that $i\alpha$ does not involve the $\pm$ sign as it appears in the phase factor, which already take opposite sites for hoppings toward different directions; and the parameter $r$ describes two different inter-cell hoppings, hence the non-reciprocities can be added in two different ways (described by $r_1$ and $r_2$ respectively), as demonstrated in Fig. \ref{fig:CL}(d) together with all other parameters.

With these extra non-Hermitian terms introduced to the model, 
a key consequence is that certain non-reciprocities along the 1D ladder may be induced to the system, and some of which lead to the NHSE. Interestingly, the NHSE does not always alters the conventional bulk-boundary correspondence in our model. These results are briefly summarized in Table \ref{table1}.
Following the conclusion in Refs. \cite{zhang2019correspondence,okuma2020topological}, the existence of the NHSE has a correspondence to a nontrivial point-gap topology of the system \cite{gong2018topological}, described by a nonzero winding of the spectrum in its Brillouin zone. 
A system free from the NHSE shall generally have its eigenenergies moving along an arclike spectrum back and forth when $k$ varies along the Brillouin zone, therefore its eigenenergies $E(k)$ are paired between $k$ and $k'$, satisfing 
$$E(k)=E(k')$$ 
with $k\neq k'$ for any given crystal momentum $k$ (except for the end points of the arclike spectrum) \cite{yao2018edge,yokomizo2019non,Lee2019anatomy,li2019geometric}.
To give a clear demonstration, we shall 
use this criterion to analyze the NHSE and topological in-gap edgemodes for different non-Hermitian parameters 
separately. 

\begin{table}
\centering
{
\begin{threeparttable}
\begin{tabular}{|c|c|c|c|}\hline
\makecell[c]{non-Hermitian \\parameter} & non-reciprocity & NHSE &\makecell[c]{conventional \\BBC}\\    \hline
$m$ & No & No & Yes\\    \hline
$r_1$ & balanced & No &Yes\\    \hline
$\alpha$& balanced & No & Yes\\    \hline
$r_2$& Yes & Yes & Partially yes$^{\ast}$\\    \hline
$\mu$ & Yes$^{\dagger}$ & Yes & No$^{\ddagger}$\\ \hline
\end{tabular}
\begin{tablenotes}
      \small
      \item[$\ast$] The conventional BBC is preserved in a certain parameter regime.
      \item[$\dagger$] The non-reciprocity appears after a rotation in the pseudospin space.
      \item[$\ddagger$] Two types of NHSE exist in this case, and only one of them breaks the conventional BBC.
    \end{tablenotes}
  \end{threeparttable}
}
\caption{Summary of the non-reciprocities along $x$ and the resulting properties of different non-Hermitian parameters. A longitudinal non-reciprocity corresponds to asymmetric hoppings along the 1D ladder, and a transverse one corresponds to that between the two sublattices.}
\label{table1}
\end{table}

\section{Without the non-Hermitian skin effect}\label{III}
In this section we will discuss the cases without the NHSE, i.e. with non-Hermiticity induced by $m$, $r_1$, or $\alpha$. 
\subsection{$\theta\rightarrow\theta+ i\alpha$}
 \begin{figure} 
\includegraphics[width=1\linewidth]{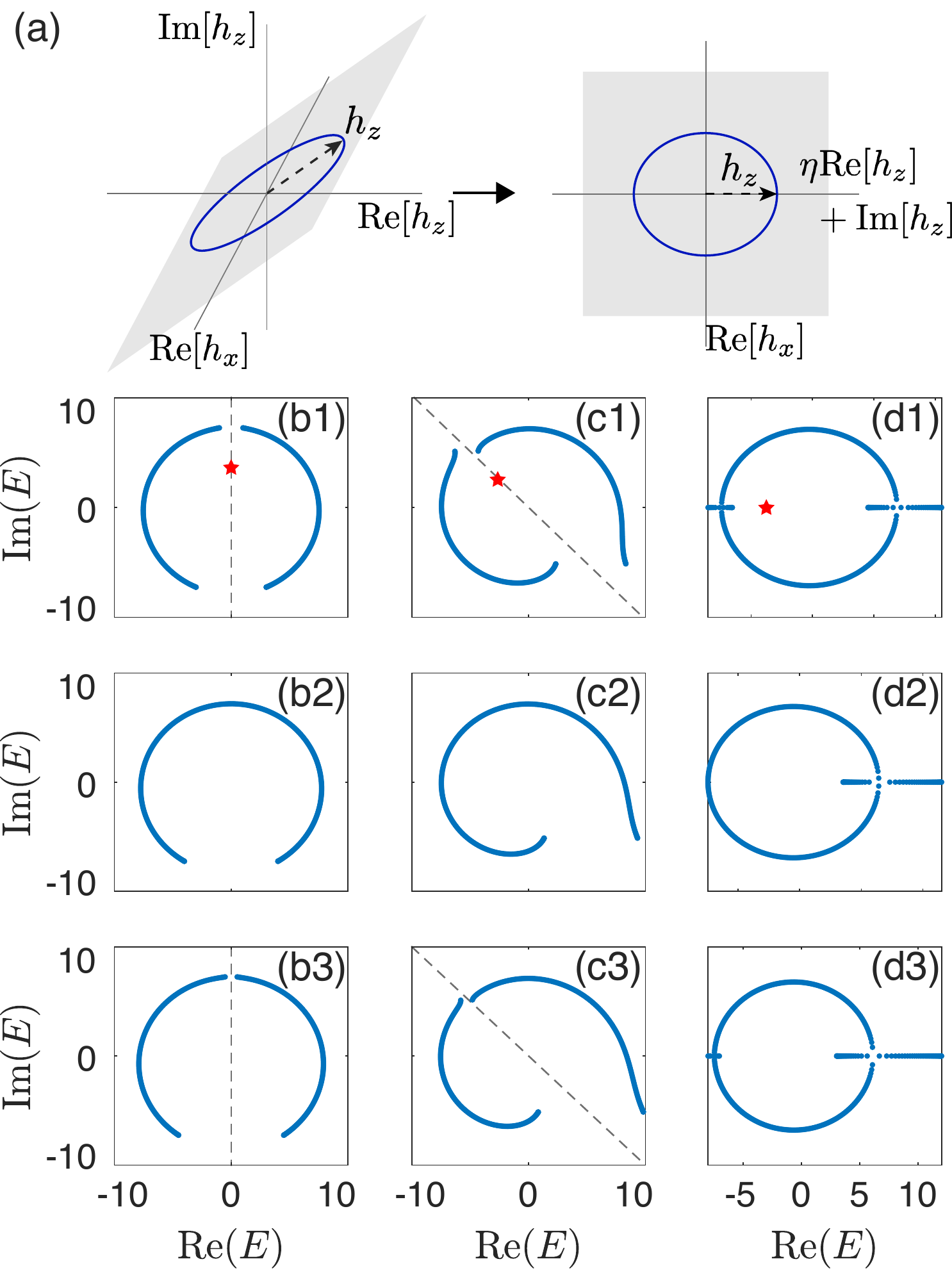}
\caption{(a) Sketches of the winding of $(h_x,h_z)$ within a 2D plane in the 4D space of $({\rm Re}[h_x],{\rm Im}[h_x],{\rm Re}[h_z],{\rm Im}[h_z])$, for the system described by Eq. \eqref{eq:H_alpha}. The axis of ${\rm Im}[h_x]$ is omitted as $h_x$ takes real value in this system.
(b) to (d)
OBC spectra of the system under different parameters. Red stars represent a pair of degenerate eigenmodes localized at opposite ends of the 1D ladder with $L=200$. 
The parameters are $\theta=\pi/2,\pi/4,0$ from (b) to (d), and $M=1,2,2.5$ from top to bottom. The middle row is at the topological transition point with $M=2r$. Other parameters are $r=1$ and $\alpha=2$. Dash lines indicate line gaps in (b) and (c), 
and in (d) the spectrum only possesses a point-gap as it encloses a finite area.
}
\label{fig:alpha}
\end{figure}

We first consider the case with an imaginary phase factor, which leads to hoppings with asymmetric  amplitudes given by $e^{\pm\alpha}$ toward right and left (left and right) for sublattice $A$ ($B$) respectively. The non-reciprocities of the two-sublattices are thus balanced and the NHSE is expected to be absent in this case.
The Hamiltonian in momentum space is given by
\begin{eqnarray}
h(k)&=&h_0\tau_0+h_z\tau_z+h_x\tau_x,\nonumber\\
h_0&=&[(e^\alpha+e^{-\alpha})\cos \theta+i(e^{-\alpha}-e^{\alpha})\sin\theta]\cos k,\nonumber\\
h_z&=&-[(e^\alpha+e^{-\alpha})\sin\theta-i(e^{-\alpha}-e^{\alpha})\cos\theta]\sin k,\nonumber\\
h_x&=&M+2r\cos k,\label{eq:H_alpha}
\end{eqnarray}
And we can see that system experiences no NHSE as it satisfies a symmetry 
\begin{eqnarray}
\sigma_xh(k)\sigma_x=h(-k), \label{eq:alpha_symmetry}
\end{eqnarray}
which means $E_\pm(k)=E_\pm(-k)$ and the eigenmodes are paired between $\pm k$. 

Now that  $h_z$ contains both sine and cosine of $\theta$, a nontrivial winding of $(h_x,h_z)$ may exist even when $\theta=0$, yet it lies in the 2D plane of (the real part of) $h_x$ and the imaginary part of $h_z$. 
More generally, $(h_x,h_z)$ can now be taken as a four-dimensional (4D) real vectors $({\rm Re}[h_x],{\rm Im}[h_x],{\rm Re}[h_z],{\rm Im}[h_z])$. Nevertheless, in our model the trajectory of this vector with $k$ varying from $0$ to $2\pi$ always lies in a 2D plane in the 4D space, as $h_x$ is real and the ratio of 
$$\eta\equiv\frac{{\rm Re}[h_z]}{{\rm Im}[h_z]}=\coth \alpha\tan \theta$$
is a $k$-independent constant [see Fig. \ref{fig:alpha}(a) for a sketch]. Consistently, we observe a pair of degenerate edge modes separated from the bulk bands when the parameters satisfy
\begin{eqnarray}
|M|<|2r|,
\end{eqnarray}
where the system possesses a nontrivial winding of $(h_x,h_z)$ in the plane shown in Fig. \ref{fig:alpha}(a). 
That is, the topological phase transition is only associated with $M$ and $r$, which is further confirmed by the transfer matrix approach in Appendix \ref{app:TM_alpha}.

Finally, we note that due to the non-Hermiticity, the eigenenergies now distributed in a 2D complex plane, hence the definitions of energy gap and in-gap eigenmodes are also richer than that for Hermitian systems.
In our model, the two bands are well-separated when $\theta\neq 0$ (or $\pi$), and a line-gap exists except for the topological transition point, as shown by the dash lines in Fig. \ref{fig:alpha}(b) and (c). 
Such a line-gap is an analog of the energy gap of Hermitian systems, which is given by a line parallel to the imaginary axis in the context of non-Hermitian systems.
In Fig. \ref{fig:alpha}(d) with $\theta=0$, the two bands always connect to each other, leading to the absence of a line-gap. 
Neverthekess, the spectrum is seen to have a point-gap, 
and a pair of degenerate edge eigenmodes appears to be separated from the bulk bands when $|M|<|2r|$, where $(h_x,h_z)$ corresponds to a nontrivial winding.

\subsection{$M\rightarrow M\pm m$ and $r\rightarrow r\pm r_1$}
Next we consider the other two non-Hermitian terms of $m$ and $r_1$ 
which enter the Hamiltonian in similar manners.
That is, they both induce even functions of $k$ to the third Pauli matrices (which is $\sigma_y$ in the Creutz ladder model), and now the Hamiltonian matrix in the momentum space reads
\begin{eqnarray}
h(k)&=&h_0\tau_0+h_z\tau_z+h_x\tau_x+h_y\tau_y,\nonumber\\
h_0&=&2\cos k\cos \theta,~~
h_z=-2\sin k\sin\theta,\nonumber\\
h_x&=&M+2r\cos k,~~h_y=i(m+2r_1\cos k). \label{eq:hk_m_r1}
\end{eqnarray}
Note that the NHSE is also expected to be absent in this scenario, as both of these two parameters do not induce a net non-reciprocity along $x$ direction: $m$ describes asymmetric hoppings between the two sublattices within the same unit cell, and $r_1$ induces opposite non-reciprocities for hoppings between sublattices $(a_x,b_{x+1})$ and sublattices $(a_x,b_{x-1})$.
Indeed, this system is free from the NHSE, as it satisfies
\begin{eqnarray}
\sigma_x h(k) \sigma_x=h^T(-k),
\end{eqnarray}
so that its PBC eigenenergies are paired between $\pm k$, i.e. $E_\pm(k)=E_\pm(-k)$.
Therefore the system obeys the conventional BBC and its topological phase transitions are given by PBC gap closing points.
To determined the phase boundaries, we rewrite the Bloch eigenenergies as
\begin{eqnarray}
E_\pm(k)=h_0\pm\sqrt{P(k)}
\end{eqnarray}
with
\begin{equation}
P(k) = (M+2r\cos{k})^2+(-2\sin{\theta}\sin{k})^2-(m+2r_1\cos{k})^2
\end{equation}
taking a real value, the square root of which gives the energy difference between the two bands at each $k$. 

\begin{figure}
\includegraphics[width = 1\linewidth]{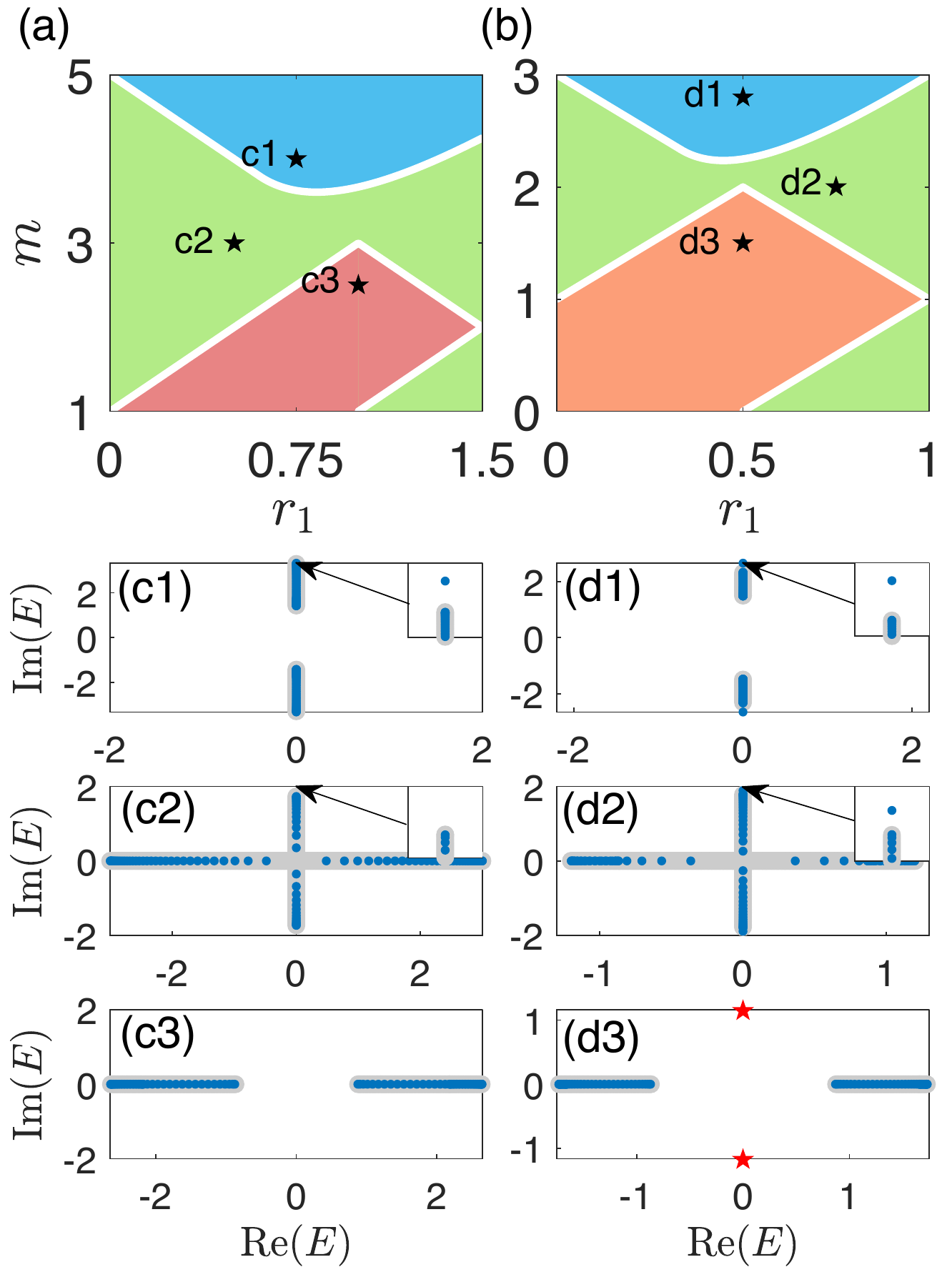}
\caption{(a) (b) Phase diagrams of the system described by Eq. \eqref{eq:hk_m_r1}, with (a) $M=3$, $r=1$, $\theta=\pi/2$ and (b) $M=r=1$, $\theta=\pi/2$.
The white solid line are obtained from conditions of Eqs. \eqref{eq:condition1_m_r1} to \eqref{eq:condition3_m_r1}.
(b) and (c) the PBC (gray) and OBC (blue dots and red stars) spectra at the black stars in (a) and (b) respectively, for a system with $L=50$ unit cells. The parameter are
(c1) $m = 4$, $r_1 = 0.75$, (c2) $m = 3$, $r_1 = 0.5$, (c3) $m = 2.5$, $r_1 = 1$, (d1) $m = 2.8$, $r_1 = 0.5$, (d2) $m = 2$, $r_1 = 0.75$, and (d3) $m = 1.5$, $r_1 = 0.5$. Red stars in (d3) are edgemodes protected by a real line-gap. 
Insets in (c) and (d) show isolated edgemodes with zero real energies, which are not topologically protected due to the absence of a real ling-gap.
}
\label{fig:phase_m_r1}
\end{figure}

We first consider the simplest case with $\theta=\pm\pi/2$, where $h_0=0$ and the two bands do not overlap. 
The system satisfies a non-Hermitian chiral symmetry \cite{kawabata2019symmetry,li2019geometric}
\begin{equation}
\sigma_yh(k)\sigma_y=-h(k),
\end{equation}
and has four topologically different phases depending on the value of $P(k)$ and the existence of in-gap edgemmodes \cite{li2019geometric}.
That is, the system has a real line-gap when
\begin{eqnarray}
P(k) > 0 ~~\forall k,\label{eq:condition1_m_r1}
\end{eqnarray}
an imaginary line-gap when
\begin{eqnarray}
P(k) < 0 ~~\forall k,\label{eq:condition2_m_r1}
\end{eqnarray}
and is gapless with the band-touching point at $k_0$ when
\begin{eqnarray}
\exists P(k)_{\rm min}<P(k_0)< P(k)_{\rm max},~P(k_0) =0.\label{eq:condition3_m_r1}
\end{eqnarray}
The topologically distinguished phases can further be identified when a real line-gap exist, manifested by the existence or absence of in-gap edgemodes with the same real energy.

In Fig. \ref{fig:phase_m_r1}(a) and (b) we illustrate the phase diagrams of our system with $\theta=\pi/2$ in different parameter regimes, where the phase boundaries are given by Eqs. \eqref{eq:condition1_m_r1}, \eqref{eq:condition2_m_r1}, and \eqref{eq:condition3_m_r1}, as detailed discussed in Appendix \ref{app:m_r1}. Typical spectra for the different phases are shown in Fig. \ref{fig:phase_m_r1}(c) and (d). 
Note that in the gapless and the imaginary-gapped phases, edgemodes may also appear apart from the PBC spectra, as shown in Fig. \ref{fig:phase_m_r1}(c1), (d1) and (d2). Nevertheless, now these edgemodes, having nonzero imaginary energies only, can continuously merge into the bulk without changing the gapped or gapless nature of the spectrum, meaning that they are not topologically protected. 
On the other hand, similar edgemodes with zero real ennergy also appear in panel (d3), where they are topologically protected by the real line-gap  \cite{li2019geometric}.

\begin{figure}
\includegraphics[width = 1\linewidth]{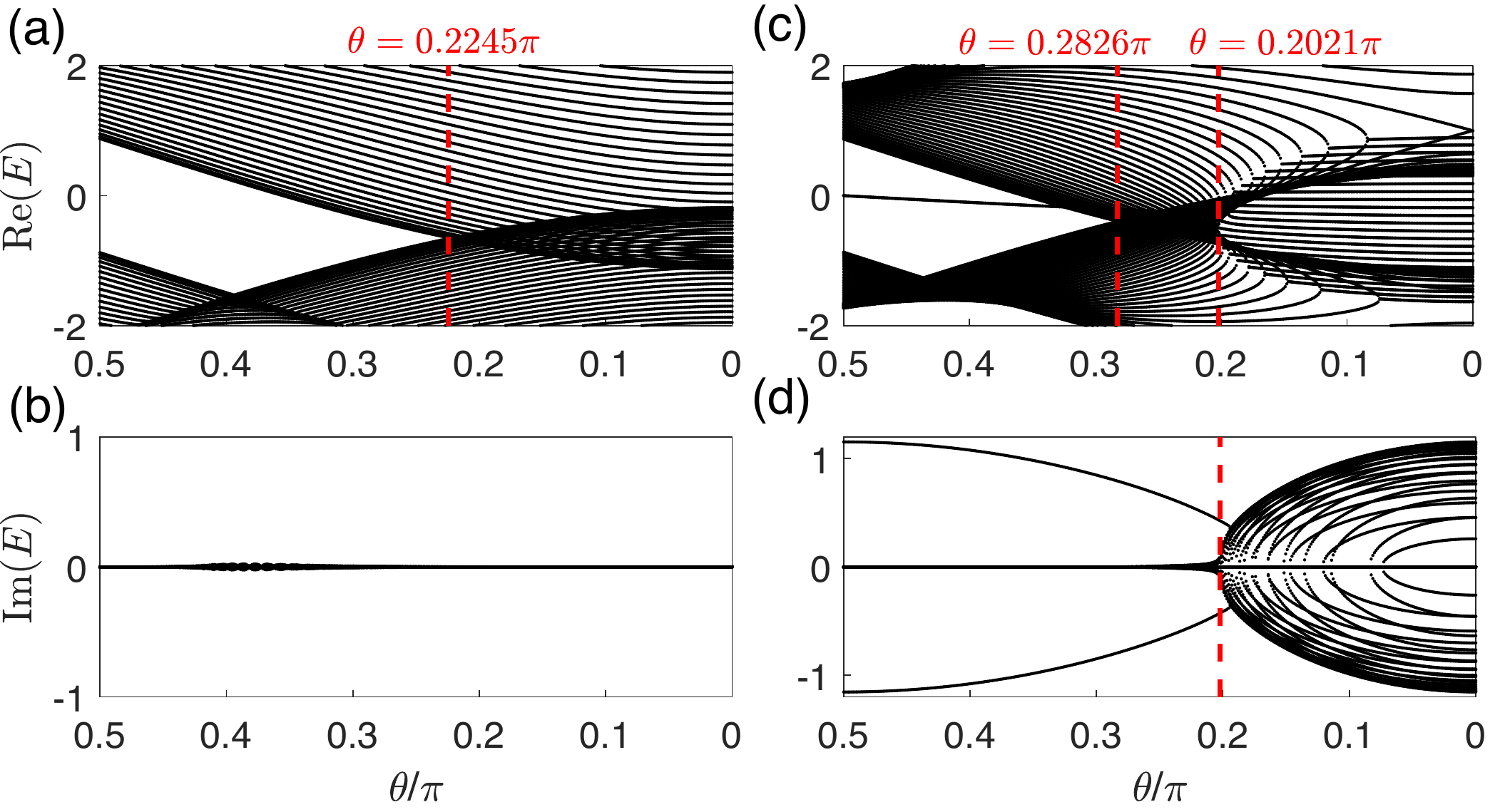}
\caption{
OBC spectra of the system described by Eq. \eqref{eq:hk_m_r1}, with $\theta$ varying from $\pi/2$ to $0$.
The parameters are $M=3$, $r=1$, $m=2.5$, $r_1=1$ for (a) and (b), the same that of Fig. \ref{fig:phase_m_r1}(c3);
and $M=r=1$, $m=1.5$, $r_1=0.5$ for (c) and (d), the same that of Fig. \ref{fig:phase_m_r1}(d3).
The red dash line in (a) and the first one in (c) are obtained From the condition of Eq. \eqref{eq:overlap_m_r1}, describing the transition from a real-gapped phase to a band-overlapping gapless phase.
Those at $\theta\approx0.2021\pi$ in (c) and (d) are obtained from the conditions of Eqs. \eqref{eq:condition1_m_r1} to \eqref{eq:condition3_m_r1},
describing the transition to a band-touching gapless phase, where some eigenenergies of the bulk eigenmodes acquire imaringary values.}
\label{fig:spectrum_m_r1_thetaNeq}
\end{figure}

For more general cases with $\theta\neq \pm\pi/2$, a nonzero but real $h_0$ shifts the real part of the eigenenergies with the same amount for each $k$. Thus another gapless phase with ${\rm Re}[E_\pm]$ overlap each other may emerges when $\theta$ is tuned away from $\pm\pi/2$, as illustrated in Fig. \ref{fig:spectrum_m_r1_thetaNeq}. The overlapping condition is given by 
\begin{eqnarray}
&&P(k) > 0 ~~\forall k,\nonumber\\
&&E_+(k)_\text{min} < E_-(k)_\text{max},\label{eq:overlap_m_r1}
\end{eqnarray}
and the corresponding phase transition point can be determined accordingly, as shown in Fig.\ref{fig:spectrum_m_r1_thetaNeq}(a) at $\theta\approx 0.22\pi$, and in Fig.\ref{fig:spectrum_m_r1_thetaNeq}(c) at $\theta\approx 0.28\pi$.
On the other hand, tuning $\theta$ also affects $P(k)$ through $h_z=-2\sin k\sin\theta$, and can lead to another transition to the gapless phase, satisfying the condition of Eq. \eqref{eq:condition3_m_r1}. Such a phase transition is shown in Fig.\ref{fig:spectrum_m_r1_thetaNeq}(c) and (d) at $\theta\approx 0.20\pi$. The critical values of these transition points are also obtained analytically (see Appendix \ref{app:m_r1}).

\section{With the non-Hermitian skin effect}\label{IV}
In contrast to the scenarios discussed above, the other two non-Hermitian parameters $r_2$ and $\mu$ lead to certain non-reciprocities along the 1D ladder, as elaborated below.
\subsection{$r\rightarrow r\pm  r_2$}
\begin{figure} 
\includegraphics[width=1\linewidth]{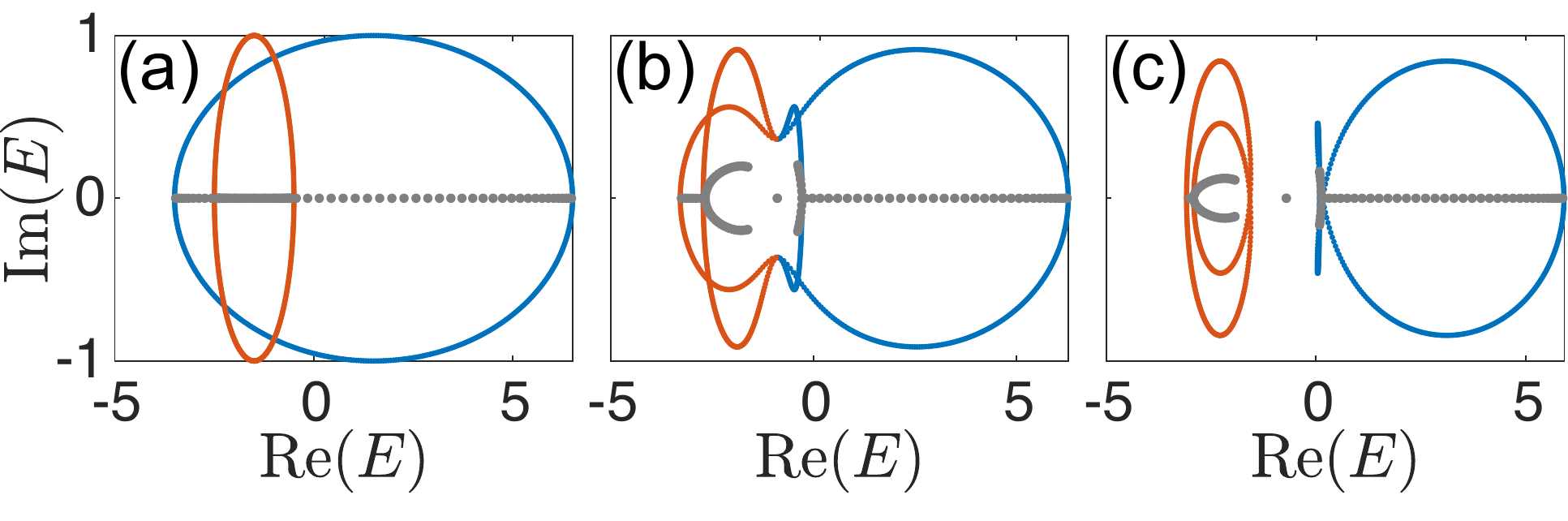}
\caption{
The PBC (red and blue) and OBC (gray) spectra, with (a) $\theta=0$, (b) $\theta=0.15\pi$, and (c) $\theta=0.25\pi$.
In panel (a), red and blue colors indicate the PBC spectra corresponding to the two decoupled sub-chains with $\sigma_x=\pm1$ respectively.
In (b) and (c), these two components are coupled, and the colors are marked regarding 
$E_+(k)$ (blue) and $E_-(k)$ (red) with ${\rm Re}[E_+(k)]>{\rm Re}[E_-(k)]$. A winding number $\nu(E^r)=2$ can be obtained for a reference energy $E^r$ enclosed by both loops in (a), or enclosed by the red one twice in (c).
Other parameters are $M=r=1.5$ and $r_2=0.5$. 
}
\label{fig:r2_NHSE}
\end{figure}

In the presence of a non-zero $r_2$, the Bloch Hamiltonian is given by
\begin{eqnarray}
h(k)&=&h_0\tau_0+h_z\tau_z+h_x\tau_x,\nonumber\\
h_0&=&2\cos k\cos \theta,~~
h_z=-2\sin k\sin\theta,\nonumber\\
h_x&=&M+2r\cos k+2ir_2\sin k. \label{eq:hk_r2}
\end{eqnarray}
When $\theta=0$ or $\pi$, the pseudospin is polarized along $\sigma_x$ direction, and the system can be decoupled 
into two sub-chains corresponding to $\sigma_x=\pm 1$ respectively, each corresponds to a single-band Hatano-Nelson model \cite{HN1996prl}. 
Explicitly, the sub-chain with $\sigma_x=1$ ($-1$) has asymmetric nearest-neighbor hopping amplitudes $t_\pm=r+1\pm r_2$ ($t_\pm=r-1\pm r_2$) toward opposite directions, and a uniform on-site potential $M$ ($-M$).
The asymmetric hoppings induce the NHSE to each sub-chain, corresponding to a loop spectrum characterized by a nonzero spectral winding number
\begin{eqnarray}
w(E^r)=\frac{1}{2\pi}\int_0^{2\pi} dk \frac{d}{dk}\arg\left[\det(h(k)-E^r)\right],\label{eq:winding}
\end{eqnarray}
with $w(E^r)$ for a reference energy $E^r$ enclosed by the spectral loop.
The full spectrum of the system form two loops in the complex plane, and gives a spectral winding number $\nu(E^r)=2$ for $E^r$ enclosed by both loops, as shown in Fig. \ref{fig:r2_NHSE}(a). 
Yet this high winding is relatively trivial as it is merely a sum of the winding numbers of two independent systems.
On the other hand, with $\theta$ tuned away from $0$ and $\pi$, a nonzero $h_z$ emerges and couples the two sub-chains. 
The two spectral loops of $\sigma_x=\pm1$ are thus restructured into two bands, which can inherit the spectral windings of both decoupled sub-chains in certain parameter regimes [see Figs. \ref{fig:r2_NHSE}(a) to (c)]. As a result, the high spectral winding is now an intrinsic property of a single band, even though the overall system possesses only nearest neighbor hoppings.

Despite the existence of a nontrivial spectral winding and the NHSE, the conventional BBC is not always violated in this system. 
To see this, we first write down the eigenenergies of the system under PBC, given by
\begin{eqnarray}
E_{\pm}&=&2\cos k\cos\theta\pm\sqrt{P(k)}
\end{eqnarray}
with $P(k)=4\sin^2 k\sin^2\theta+(M+2r\cos k+2ir_2\sin k)^2$.
The touching of the two bands requires 
${\rm Re}[P(k)]={\rm Im}[P(k)]=0$,
namely
\begin{eqnarray}
(M+2r\cos k)^2+4\sin^2 k(\sin^2 \theta-r_2^2)&=&0,\label{eq:gapless1}\\
4r_2\sin k(M+2r\cos k)&=&0.\label{eq:gapless2}
\end{eqnarray}
These conditions are satisfied under two parameter regimes:

(i) a normal degenerate point (DP) of the two bands arises at $k=\pi$ or $0$ when 
\begin{eqnarray}
M+2r\cos k=\sin k=0,~M=\pm2r,\label{eq:r2_con1}
\end{eqnarray}
and 

(ii) two exceptional degenerate points (EPs) arise at $\cos k=-M/2r$ when
\begin{eqnarray}
M+2r\cos k=0,~|r_2|=|\sin\theta|,~|M|<|2r|.\label{eq:r2_con2}
\end{eqnarray}

\begin{figure} 
\includegraphics[width=1\linewidth]{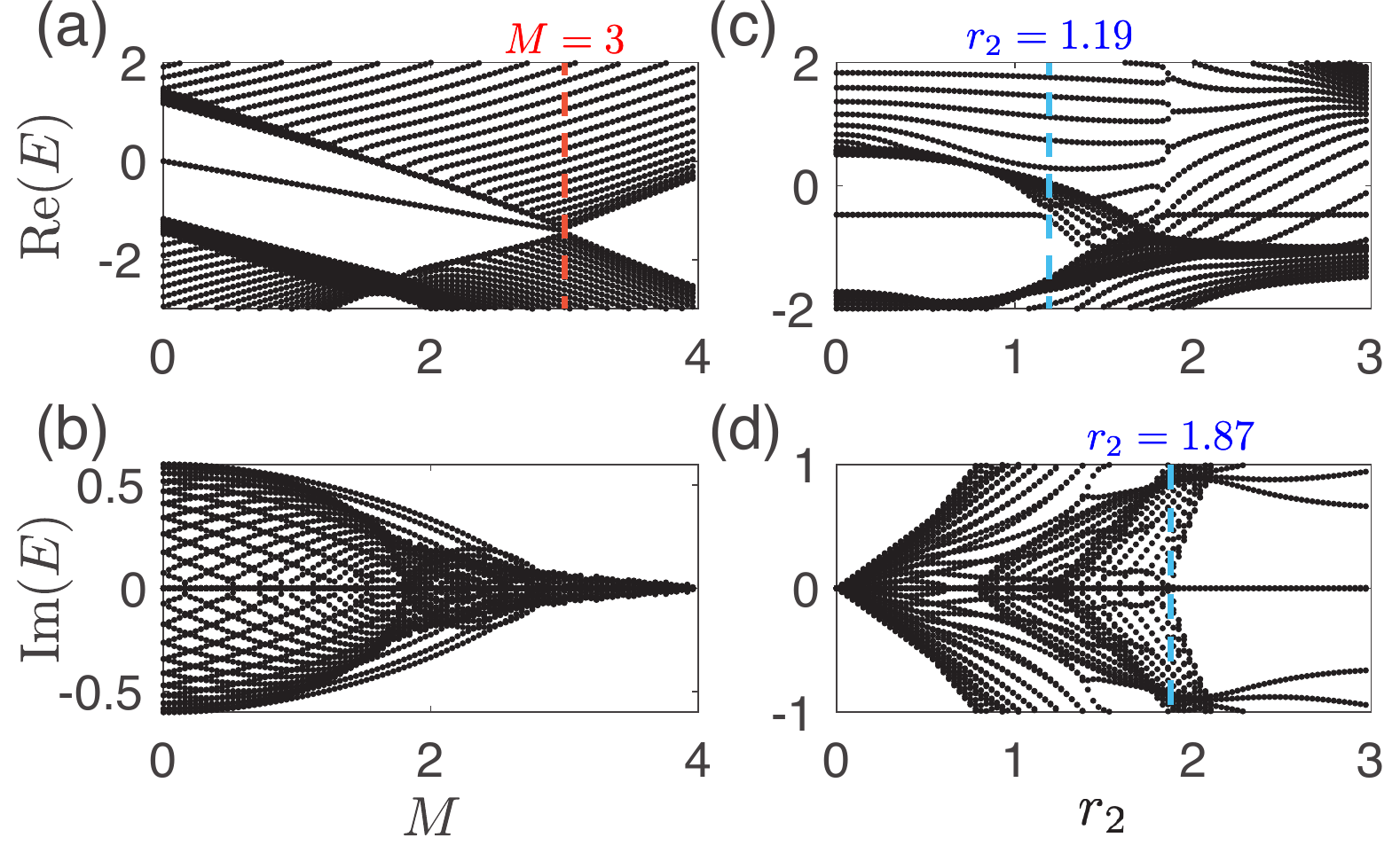}
\caption{
OBC spectra of the system described by Eq. \eqref{eq:hk_r2} with $r=1.5$, $\theta=\pi/4$.
(a) and (b) the real and imaginary parts of the eigenenergies with $r_2=0.5$, as a function of $M$. The conventional BBC preserves as $|r_2|<|\sin\theta|$. The red dash line indicates the topological phase transition at $M=2r=3$, which is also the gap closing condition of Eq. \eqref{eq:r2_con1} for the PBC spectrum. 
(c) and (d) the real and imaginary parts of the eigenenergies with $M=1$, as a function of $r_2$. The conventional BBC may be broken when $r_2>\sin\theta=\sqrt{2}/2$. The dash lines indicate two transitions between topological gapped phases and a gapless phase, where the critical values of $r_2$ are obtained numerically (also see Fig. \ref{fig:r2_phases}) .
}
\label{fig:r2_spectrum}
\end{figure}

When $|r_2|<|\sin\theta|$, the PBC spectrum is seen to form two separated loops (e.g. as in Fig. \ref{fig:r2_NHSE}(c)), and the line-gap between them may close only at a DP satisfying the above condition (i). 
In such a scenario, the topological phase transition under the OBC are also seen to occurs at the DP with $M=\pm 2r$, as shown in Fig. \ref{fig:r2_spectrum}(a) and (b). This observation suggests that the conventional BBC still holds in this parameter regime.
The reason is that the spectral winding topology requires the OBC arc-like spectrum to fall within the PBC loop-like one \cite{zhang2019correspondence,okuma2020topological}, so that the gap-closing conditions for them may be different from each other only after the system goes through an exceptional PBC gap-closing, as further explained in Appendix \ref{app:BBC_NHSE} .


With the amplitude of $r_2$ increases, two EPs satisfying the above condition (ii) when $|r_2|=|\sin\theta|$, and the two loops connect each other and reform two overlapping loops (e.g. as in Fig. \ref{fig:r2_NHSE}(a) and (b)). As now a line-gap is absent for the PBC spectrum, the two OBC bands may go through a gap-closing in the interior of the PBC loops, without generating either a DP or an EP to the PBC spectrum (see Appendix \ref{app:BBC_NHSE}). 
 
In Fig. \ref{fig:r2_phases}(a) we illustrate a phase diagram of the system, where four different phases are seen from our numerical results [Fig. \ref{fig:r2_phases}(b) to (e)]: two topologically nontrivial phases with in-gap edge modes [Fig. \ref{fig:r2_phases}(b) and (e)], a trivial gapped phase [Fig. \ref{fig:r2_phases}(d)], and a gapless phase [Fig. \ref{fig:r2_phases}(c)].
It is seen that when $|r_2|>|\sin\theta|$, the phase boundaries, including that between two topologically different gapped phases of panels (b) and (e), cannot be predicted by the PBC gap closing (white solid lines).

Note that the dash lines in Fig. \ref{fig:r2_phases}(a) are read out by comparing the spacing between different eigenenergies from the numerical results of the energy spectrum [e.g. as in Fig. \ref{fig:r2_spectrum}(c) and (d)]. 
These boundaries may not be accurate ones, as topological properties of a non-Hermitian system can also depend on the system's size \cite{li2020critical}, and numerical results of non-Hermitian Hamiltonians suffer from strong computational error for large systems \cite{yang2020non}. Furthermore, the criterion for identifying a gap also become blurred, especially when the system is near the phase boundaries, as now the eigenenergies take complex values.

\begin{figure} 
\includegraphics[width=1\linewidth]{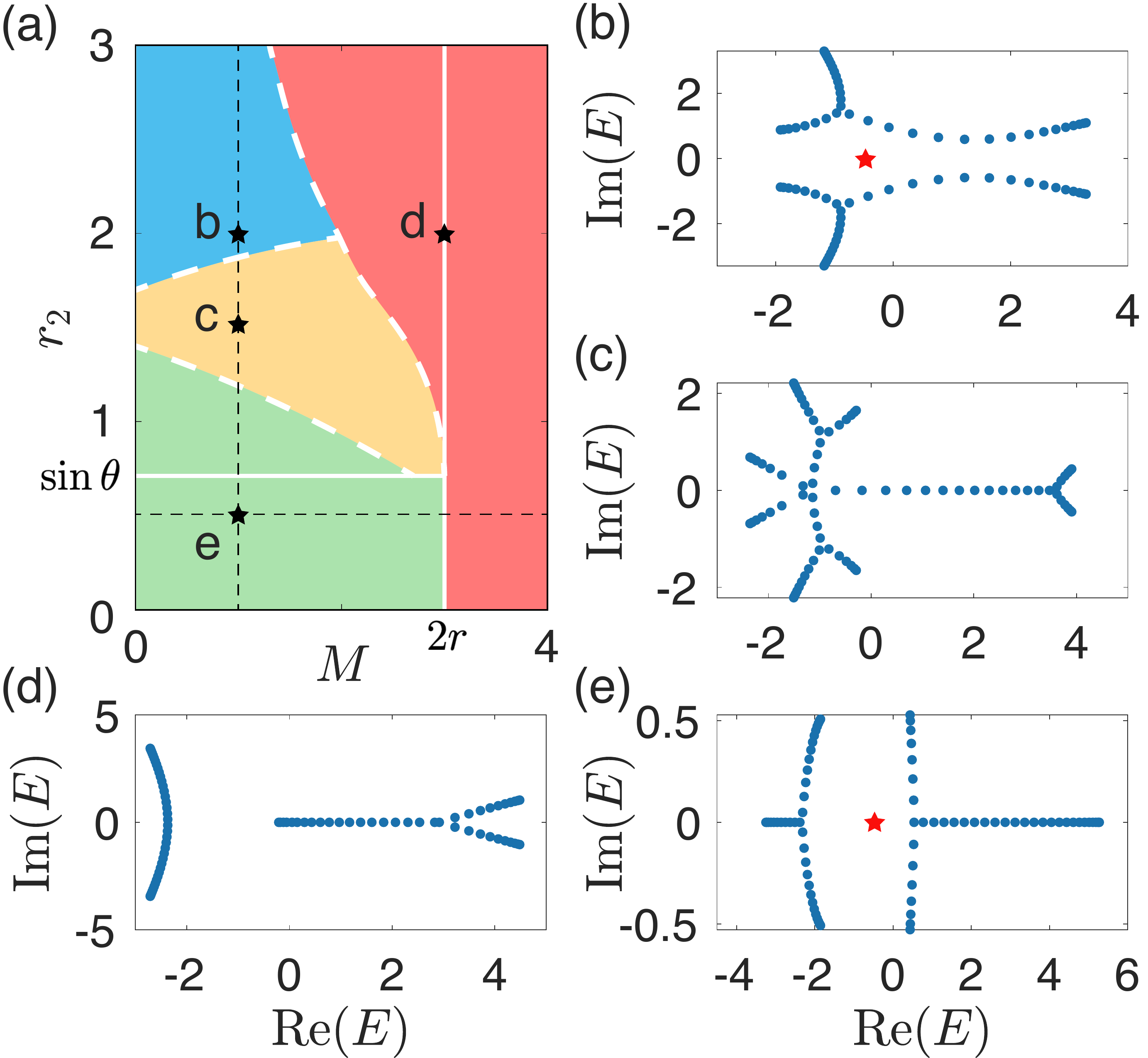}
\caption{
(a) A phase diagram of the system described by Eq. \eqref{eq:hk_r2}, with $r=1.5$, $\theta=\pi/4$, and $L=40$. The white solid lines are obtained from the conditions of Eqs. \eqref{eq:r2_con1} and \eqref{eq:r2_con2}, and the white dash lines are obtained numerically. The two black dash lines represent the parameter regions for the numerical results in Fig. \ref{fig:r2_spectrum}(a) (b) and (c) (d) respectively.
(b) to (d) OBC spectra corresponding to the four black stars in (a), with (b) $M=1$, $r_2=0.5$; (c) $M=1$, $r_2=1.5$; (d) $M=1$, $r_2=2$; and (e) $M=3$, $r_2=2$. Red stars are the topological in-gap eigenmodes.
}
\label{fig:r2_phases}
\end{figure}

\subsection{Imaginary on-site potentials}
Last but not least, we consider the case with sublattice-dependent imaginary on-site potential term $H_\mu$ adding to the Hamiltonian, 
\begin{eqnarray}
H_\mu=i\mu_a\hat{a}_x^\dagger\hat{a}_x+i\mu_b\hat{b}_x^\dagger\hat{b}_x.
\end{eqnarray}
Nonzero $\mu_{a,b}$ represent unbalance on-site dissipations on the two sublattices, which may also induce the NHSE under certain circumstances \cite{song2019non,li2020topological,yi2020nonH}.
Here we further choose $\mu_b=-\mu_a=\mu$, as any other choice is equivalent to it upon a uniform shifting of the spectrum along the imaginary axis. The overall Bloch Hamiltonian with the on-site dissipations now reads
\begin{eqnarray}
h(k)&=&h_0\tau_0+h_z\tau_z+h_x\tau_x,\nonumber\\
h_0&=&2\cos k\cos \theta,~~
h_z=-2\sin k\sin\theta-i\mu,\nonumber\\
h_x&=&M+2r\cos k.\label{eq:hk_mu}
\end{eqnarray}

Unlike the non-Hermiticity discussed in previous sections, the imaginary on-site potential $\mu$ does not directly induce non-reciprocity (i.e. asymmetric hoppings) to the system. Yet with the rotation of the pseudospin in Eq. \eqref{eq:rotation}, the parameters $M$ and $\mu$ form asymmetric intra-cell hoppings as in the non-Hermitian SSH model, and induce nontrivial spectral winding and the NHSE to the system. 
In other words, a nonzero $\mu$ induces a ``hidden" non-reicprocity to the Creutz ladder, and leads to the NHSE.
Consequently, conventional topological properties associated with in-gap edge modes shall also be affected by the NHSE \cite{yao2018edge}.
In the following of these section we shall first give an analysis of the NHSE in this model, then discuss its topological in-gap edge modes.


\subsubsection{the competition between two types of NHSE}
\begin{figure*} 
\includegraphics[width=1\linewidth]{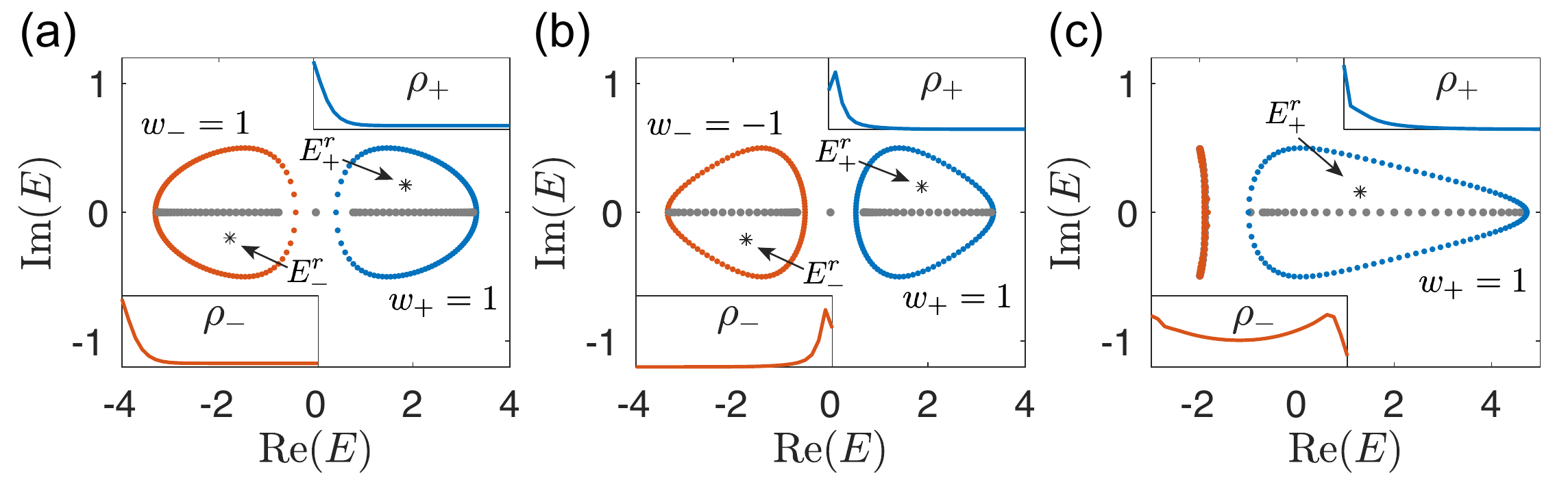}
\caption{
The PBC (red and blue) and OBC (gray) spectra, and the summed population $\rho_{\pm}(x)=\sum_{n\in\pm}[|\psi_{x,n}^A|^2+|\psi_{x,n}^B|^2]$ for each of the the two bands, where $\psi_{x,n}^{\beta}$ is the amplitude of the $n$th eigenmode on $\beta$ sublattice of the $x$th unit cell, and the summation runs over all the eigenmodes
of one of the two bands (``$+$" for the band marked blue and ``$-$" for that marked red).
The spectral winding numbers $w_\pm$ defined for the chosen reference energies $E^r_\pm$ (black stars) are indicated in the figures.
Note that the maximum value of $\rho_{-}(x)$ in (c) is $\sim 1.5$, and that of the rest are $\sim 10$ (no shown in the figure). 
The parameters are $r=1$, $\mu=0.5$, $L=30$, with (a) $M=1.35$, $\theta=\pi/2$, (b) $M=0$, $\theta=\pi/4$,  and (c) $M=1.35$, $\theta=\pi/4$.}
\label{fig:mu_NHSE}
\end{figure*}
To analyze the NHSE in this model, we shall first consider two special parameter regimes. The first one is with $\theta=\pi/2$ and $r=\sin\theta=1$.
Under such conditions, the Hamiltonian can be exactly mapped to the well-studied non-Hermitian SSH model \cite{Yin2018nonHermitian,yao2018edge,Lee2019anatomy}, with $M\pm \mu$ and $2r=2$ being the strengths of the non-reciprocal intracell and reciprocal intercell hoppings.
Thus a {\it uniform} NHSE emerges with the inverse localization length of all the skin modes given by \cite{yao2018edge}
\begin{eqnarray}
\kappa_u=\ln\sqrt{\left|\frac{M+\mu}{M-\mu}\right|},\label{eq:uNHSE}
\end{eqnarray}
i.e. eigenmodes of both bands have the same distribution [see Fig. \ref{fig:mu_NHSE}(a)].
With the chosen parameters, the PBC spectrum of $h(k)$ forms two loops for the two bands, leading to nonzero spectral winding numbers
$$w_\pm:=w(E^r_\pm)$$
defined for the two bands following Eq. \eqref{eq:winding},
with $E^r_\pm$ the reference energies chosen to be enclosed by the ``$+$" and ``$-$"bands respectively, where ``$+$" represents the band with larger ${\rm Re}[E]$.
Following the non-Bloch band theory, now the OBC system, together with its possible topological edge modes, is described by the non-Bloch Hamiltonian $h(k+i\kappa_u)$, where $k+i\kappa_u$ with $k$ varying from $0$ to $2\pi$ is dubbed as the generalized Brillouin zone (GBZ) \cite{yao2018edge,yokomizo2019non}. A key property of the GBZ is that it must give a line-shape spectrum (as that under OBC) with two-fold degeneracy (except for the end points of the lines) \cite{zhang2019correspondence}.
In the above scenario this condition is satisfied with
$$E_{u,\pm}(k+i\kappa_u)=E_{u,\pm}(-k+i\kappa_u),$$
where $E_{u,\pm}(k+i\kappa_u)$ are the eigenenergies of the Hamiltonian of Eq. \eqref{eq:hk_mu} in the GBZ, with $\theta=\pi/2$ and $r=1$.

Next we consider another parameter regime with $M=0$. The inverse localization length defined in Eq. \eqref{eq:uNHSE} becomes $\kappa_u=0$, meaning the absence of the uniform NHSE. However, the PBC spectrum remains as loops, and the OBC system are seen to exhibit a {\it bidirectional} NHSE, namely the eigenmodes of the two bands accumulate to opposite ends of the 1D ladder, as shown in Fig. \ref{fig:mu_NHSE}(b). 
The spectral winding numbers also satisfy $w_+=-w_-$ for reference energies enclosed by different bands.
The corresponding inverse localization length $\kappa_b$ is found to be $k$-dependent from our numerical computation (results not shown), and is difficult to obtain analytically. Nevertheless, it is straightforward to see that the eigenenergies in this parameter regime satisfy
\begin{eqnarray}
E_{b,+}(k+i\kappa_b)=-[E_{b,-}(k+\pi-i\kappa_b)]^*.\label{eq:E_symmetry_mu}
\end{eqnarray}
Assuming the inverse localization length is given by $\kappa_b^{\pm}(k)$ for the two bands, the OBC eigenenergies must have paired $k$ and $k'$ satisfying 
$$E_{\pm}^b(k+i\kappa_b^{\pm}(k))=E_{\pm}^b(k'+i\kappa_b^{\pm}(k)).$$ 
The exact relation between $k$ and $k'$ is unknown, yet from Eq. \eqref{eq:E_symmetry_mu} we will also have 
$$E_{-}^b(k+\pi-i\kappa_b^+(k))=E_{-}^b(k'+\pi-i\kappa_b^+(k)).$$ 
Combining the above two equations, we reach the conclusion that
$$\kappa^+_b(k)=-\kappa^-_b(k+\pi),$$
i.e. the eigenmodes associated with $k$ and $k+\pi$ have opposite inverse localization lengths for the two bands, thus they shall accumulate at different ends of the system under the OBC.

\begin{figure} 
\includegraphics[width=1\linewidth]{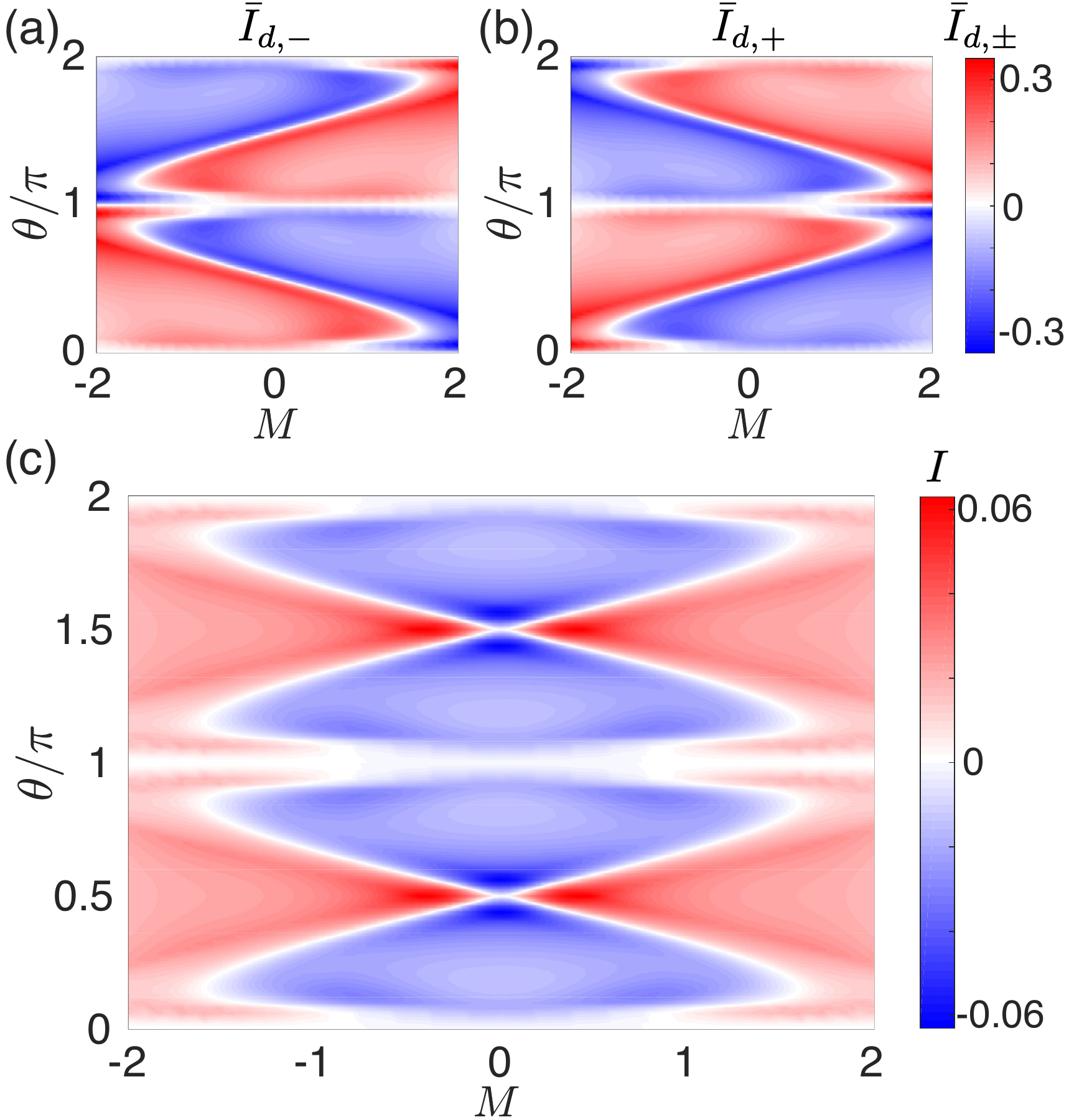}
\caption{(a) (b) The average $\bar{I}_{d,\pm}$ dIPR defined in Eq. \eqref{eq:a_dIPR} for the two bands respectively, versus the parameters $\theta$ and $M$. (c) The product of $\bar{I}_{d,+}$ and $\bar{I}_{d,-}$, within the same parameter regime as that of (a) and (b). Positive (negative) $I$ indicates that the system is dominated by the uniform (bidirection) NHSE, namely the two bands tend to localize at the same (opposite) ends of the 1D ladder.
Other parameters are $r=1$, $\mu=0.5$, and $L=30$.}
\label{fig:mu_NHSE_IPR}
\end{figure}

Now that we have two types of NHSE, the competition between them is expected in a more general parameter regime. 
Following previous discussion, the two types of NHSE shall induce opposite localizing directions to only one of the two bands, which is the one labeled with ``$-$" for the parameters we choose in  Fig. \ref{fig:mu_NHSE}(c). Specifically, when the absolute value of the inverse localization lengths of them are approximately equivalent, the ``$-$" band shall have almost identical PBC and OBC spectrum, with its eigenmodes experiencing only a very weak skin localization, as shown in  Fig. \ref{fig:mu_NHSE}(c). It also indicates that the total inverse localization length $\kappa(k)$, with contributions from both $\kappa_u$ and $\kappa_b(k)$, is not only $k$-dependent, but also takes different values for different bands.

To describe the competition between the two types of NHSE and give a phase diagram regarding which of them dominates in the system, we define a directed inverse participation ratio (dIPR) as
\begin{eqnarray}
I_{d,n}=\sum_x(x-x_c)(|\psi_{x,n}^A|^4+|\psi_{x,n}^B|^4)/[(L-1)/2],\nonumber\\
\end{eqnarray}
with $\psi_{x,n}^{\beta}$ the amplitude of the $n$th eigenmode on $\beta$ sublattice of the $x$th unit cell, and
$x_c=(L+1)/2$ being the center of the system. By definition, $I_{d,n}$ takes positive (negative)
values with larger amplitudes for eigenmodes with stronger accumulation toward $x=L(x=0)$, and $I_{d,n}=0$ for a perfectly extended eigenmode (or one with spatially balanced localization). We then further defined average dIPRs as
\begin{eqnarray}
\bar{I}_{d,\pm}=\sum_{n\in\pm} I_{d,n}/N,\label{eq:a_dIPR}
\end{eqnarray}
which describe the average localizations and their directions of the two band labeled by ``$\pm$". Here the summation runs over all eigenmodes belonging to the corresponding energy band, $N=L-1$ when the system hosts in-gap topological edge modes, and $N=L$ when it has not. We can see from Figs. \ref{fig:mu_NHSE_IPR}(a) and (b) that the two bands localize toward different directions in different parameter regimes.
Finally, in Fig. \ref{fig:mu_NHSE_IPR}(c), we illustrate the value of $I\equiv\bar{I}_{d,+}\times\bar{I}_{d,-}$, which takes positive value when the two bands share the same direction of the skin localization, i.e. the system is dominated by the uniform NHSE (the red region); and negative value when the two bands mostly localized at different ends of the system, i.e.  the system is dominated by the bidirectional NHSE (the blue region).

\subsubsection{Topological edge modes}
\begin{figure}
\includegraphics[width = 1\linewidth]{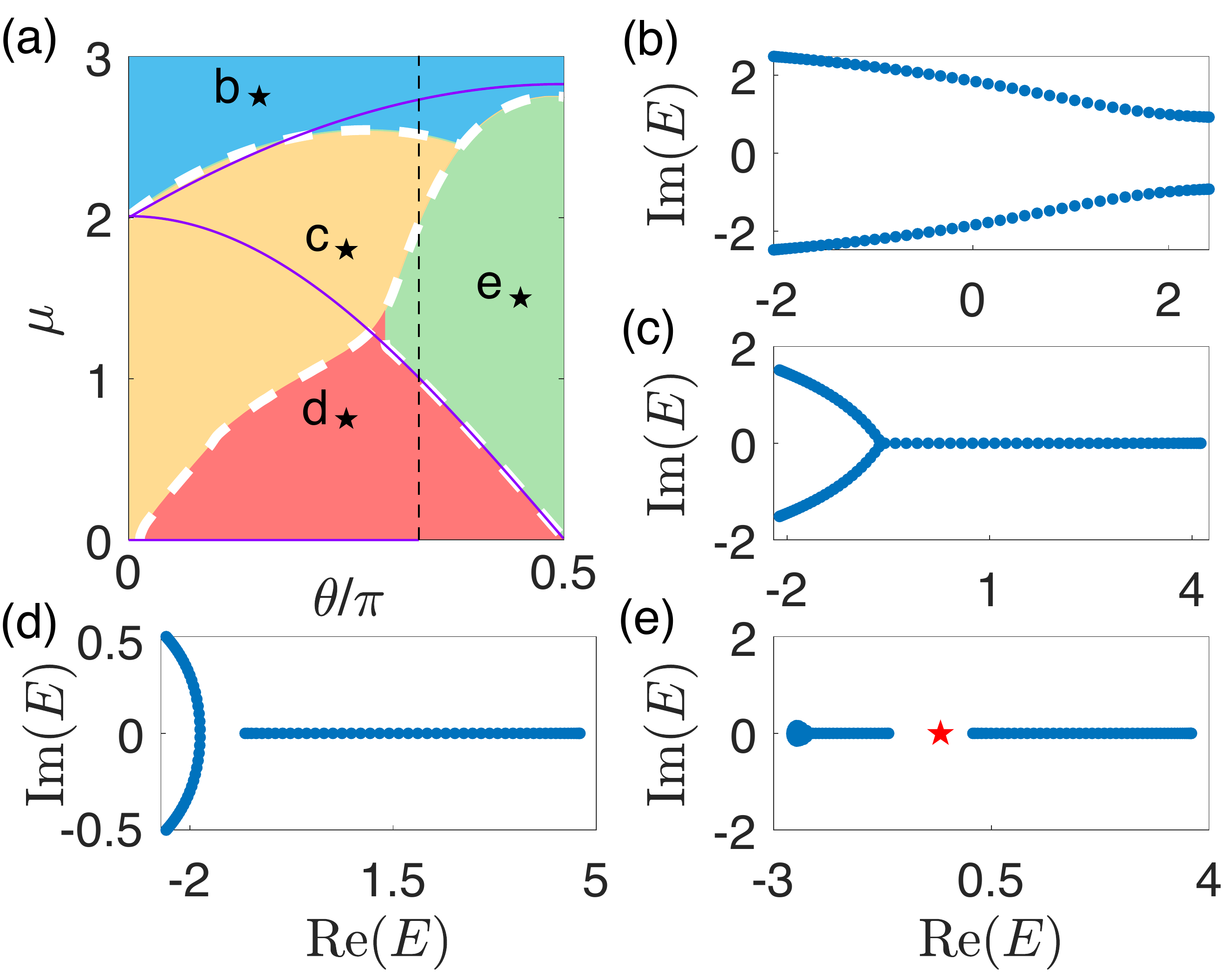}
\caption{
(a) A phase diagram of  the system described by Eq. \eqref{eq:hk_mu}, with $r = \sin{\theta}$, $M=2$, and $L=50$. 
The phase boundaries (white dashed lines) are obtained numerically, and the two purple lines are given by Eq. \eqref{eq:edge_r=sin}.
(b) to (e) the  OBC spectra corresponding to the four black stars in (a), with (b) $\theta=0.15\pi$, $\mu=2.75$; (c) $\theta=0.25\pi$, $\mu=1.8$; (d) $\theta=0.25$, $\mu=0.75$; and (e) $\theta=0.45$, $\mu=1.5$. Red stars are the topological in-gap eigenmodes.
}
\label{fig:phase_r=sin}
\end{figure}

Next we shall consider the topological in-gap edge modes and associated topological phase transitions in this model.
As in the previous subsection, we begin with the simplest case with $$\theta = \frac{\pi}{2}, ~r=\sin{\theta}=1,$$ where the system is an analog of the non-Hermitian SSH model and is analytically solvable \cite{yao2018edge}. 
Following non-Bloch band theory, the OBC topological properties of this system are described by the non-Bloch Hamiltonian $h(k+i\kappa_u)$, with $\kappa_u$ given by Eq. \eqref{eq:uNHSE} describes the locality of the skin modes under OBC. 
The system is topologically nontrivial with a pair of in-gap topological edgemodes when \cite{yao2018edge}
\begin{equation}
	\sqrt{|M^2-\mu^2|} <|2r|=2,\label{eq:edge_r=sin=1}
\end{equation}
and we can obtain the phase diagram in Fig.[\ref{fig:phase_r=sin}](a). The other panels in this figure illustrate typical OBC spectra of the three different phases, including a topologically nontrivial phases in (c), and two topologically trivial phases with real and imaginary line-gaps in (b) and (d) respectively.

\begin{figure}
\includegraphics[width = 1\linewidth]{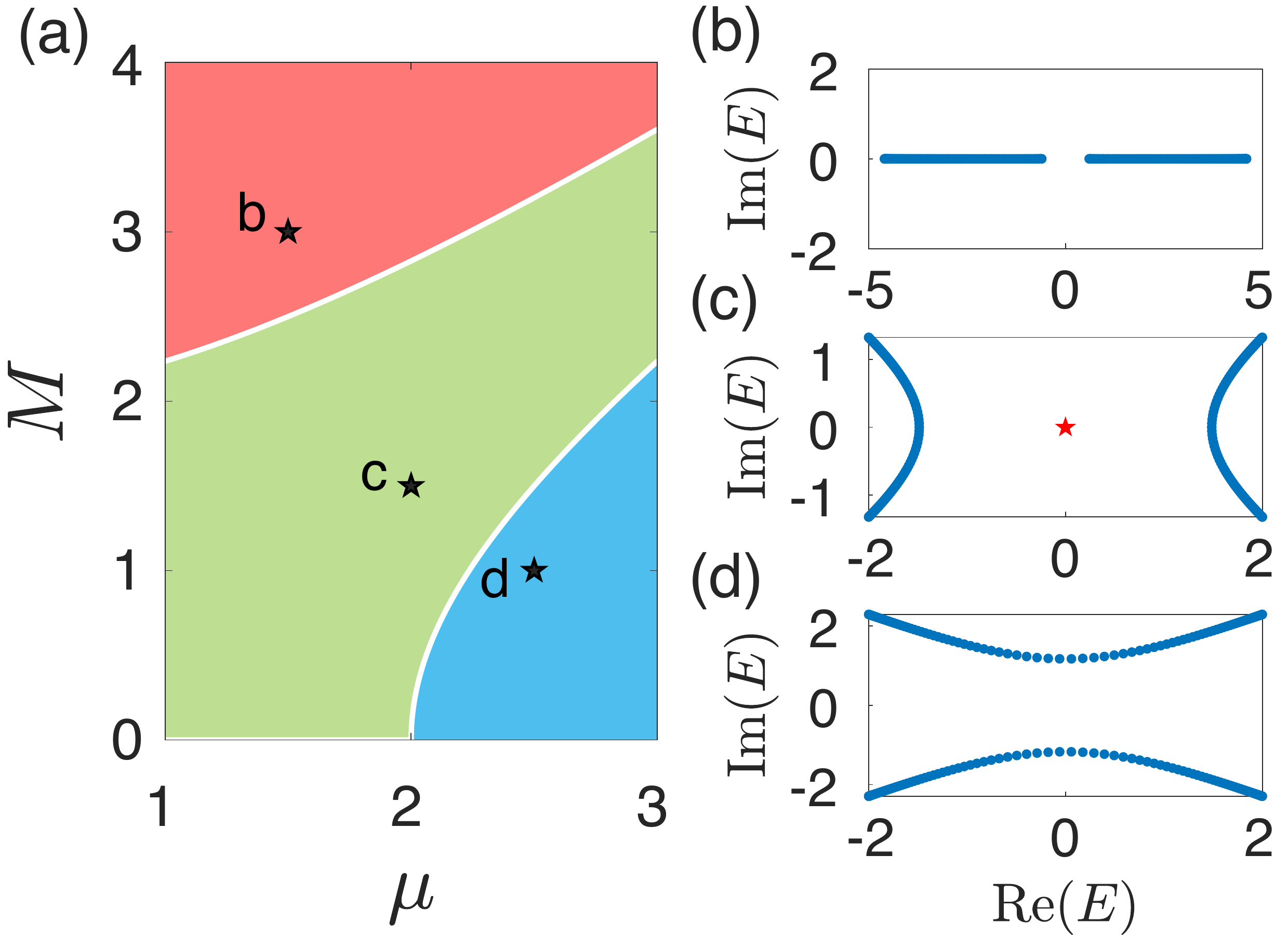}
\caption{(a) A phase diagram of  the system described by Eq. \eqref{eq:hk_mu}, with $\theta=\pi/2$, $r = \sin{\theta} = 1$, and $L=100$. 
The phase boundaries (white solid lines) are given by Eq. \eqref{eq:edge_r=sin=1}.
(b) to (d) the  OBC spectra corresponding to the three black stars in (a), with (b) $M=3$, $\mu=1.5$; (c) $M=1.5$, $\mu=2$; and (d) $M=1$, $\mu=2.5$. Red stars are the topological in-gap eigenmodes.}
\label{fig:phase_r=sin=1}
\end{figure}

\begin{figure}
\includegraphics[width = 1\linewidth]{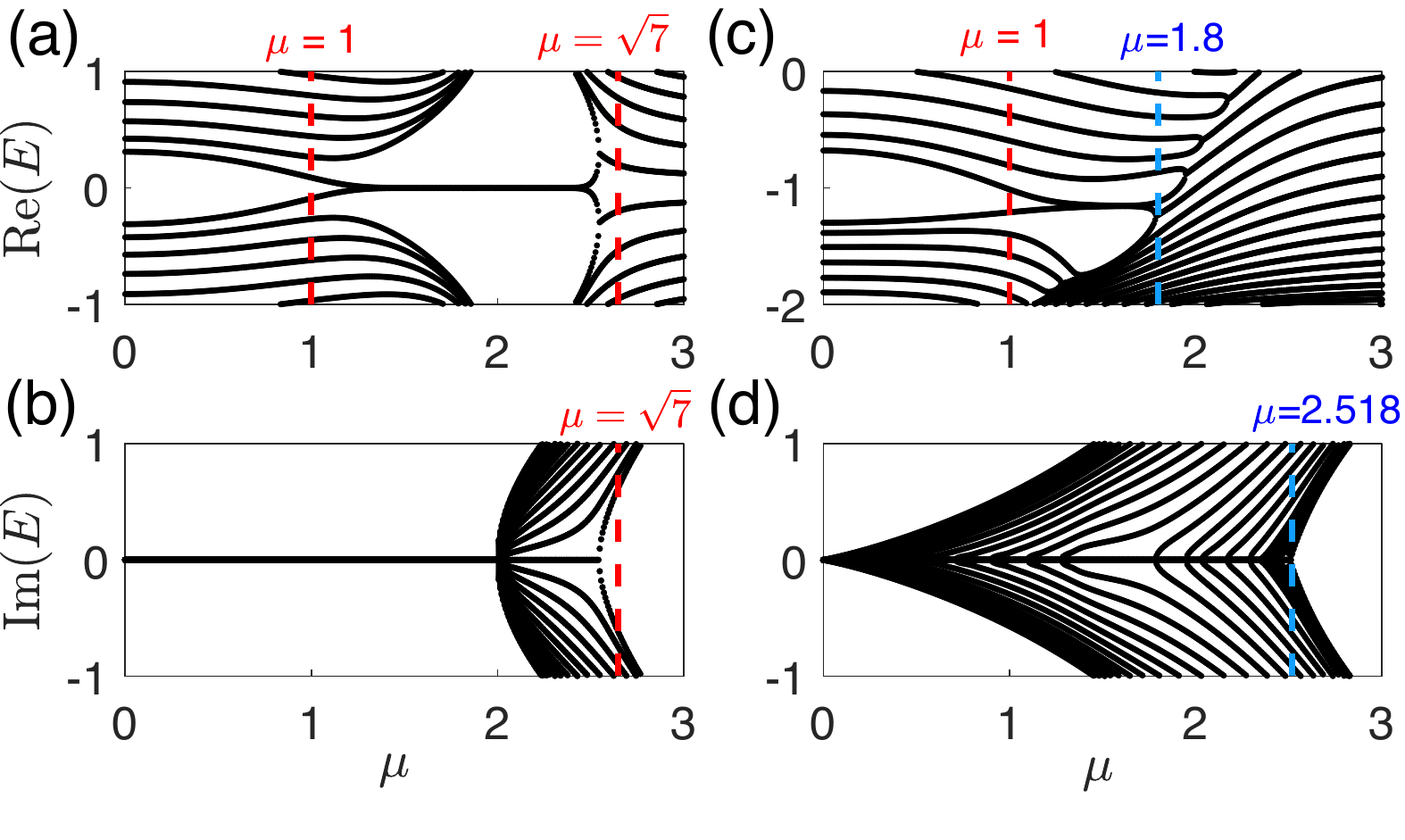}
\caption{The energy spectrum of $h_0 \neq 0$(Top) and $h_0 = 0$(Bottom). For this condition $h_0 = 0$, the red and blue dash line is the topological phase transition point ($\mu = \sqrt{|4\sin^2{\theta}-M^2|}$, $\mu = 1$ and $\mu = \sqrt{7}$). For condition $h_0\neq 0 $, the topological phase transition point and the point of imaginary spectrum starting to separate is indicated by red dash line and blue dash line, respectively. The parameter are system size $L = 30$ $\theta = \pi/3$, $M = 2$.}
\label{fig:spectrum}
\end{figure}

With $\theta$ diverges from $\pm\pi/2$, a nonzero $h_0$ term emerges and leads to more complicated phase transitions in the system. For instance, in Hermitian systems, the $h_0$ term may induce an extra gapless phase with the two bands overlapping with each other,
as shown in Fig. \ref{fig:CL}(c). However, it does not change the topology of each ``band", which is given by the eigenenergies continuously changed with the momentum $k$.
In non-Hermitian systems, as $h_0$ only shifts the two PBC bands simultaneously and does not induce any degeneracies (normal or exceptional) between them, it is also expected to not change the BBC, either a conventional one, or a non-Bloch one raised from some other terms as in the current case.
That is, the topological properties shall be described solely by the Hamiltonian matrix $h_\sigma(k):=h(k)-h_0$, which is still an analog of the non-Hermitian SSH model by setting 
$$r = \sin{\theta},$$
and a topologically nontrivial phase of $h_\sigma(k)$ occurs when
\begin{equation}
	\sqrt{|M^2-\mu^2|} <|2r|=|2\sin{\theta}|\label{eq:edge_r=sin}.
\end{equation}
As shown in Fig. \ref{fig:spectrum}(a) and (b), the above condition predicts the gap closing points of the corresponding OBC system of $h_\sigma$, which are at $\mu=1$ and $\sqrt{7}$ for the chosen parameters, and a pair of in-gap eigenmodes emerge between these two critical values.

With $h_0$ taken into account, the non-Bloch Hamiltonian is no longer given by $h(k+i\kappa_u)$, as $h_0$ is $k$-dependent and shall affects the GBZ solution of $\kappa(k)$. As mentioned previously, the value of $\kappa(k)$ dependents on both the momentum $k$ and which energy band is considered, and an analytical solution does not generally exist. 
Nevertheless, in Fig. \ref{fig:spectrum}(c) and (d) we illustrate the real and imaginary parts of the eigenenergies with the same parameters are in (a) and (b), and it is seen that the topological phase transition at $\mu=1$ is not affected by $h_0$. However, the system goes into a gapless phase at $\mu\approx1.8$, before the second topological transition point at $\mu=\sqrt{7}$. With further increasing $\mu$, an imaginary line-gap opens at $\mu\approx2.52<\sqrt{7}$, seemingly suggests that the second topological transition is absent. Nevertheless, we note that this inconsistence is due to the strong finite size effect in non-Hermitian systems, where topological transition may even occurs when changing the system's size \cite{li2020critical}. As a matter of fact, the imaginary line-gap also emerges when $\mu<\sqrt{7}$ in Fig. \ref{fig:spectrum}(b), meaning that the inconsistency is not because of the presence of $h_0$.

By further comparing the spacing between different eigenenergies, we numerically obtain the phase boundaries between the four different phases, including the three with $\theta=\pi/2$ in Fig. \ref{fig:phase_r=sin=1}, and an extra gapless phase due to the presence of $h_0$, as shown in Fig. \ref{fig:phase_r=sin}.
It is seen that the phase boundary (white dash lines) between the two topologically different phases with a real line-gap [Fig. \ref{fig:phase_r=sin}(d) and (e)] coincides with the topological phase boundary given by Eq. \eqref{eq:edge_r=sin} (purple lines), which further confirms that the non-Bloch BBC associated with $\kappa_u$ is not further modified by $h_0$. On the other hand, in the region close to $\theta=\pi/2$ and $\mu=3$, the system goes into the regime with an imaginary line-gap before approaching the analytically predicted topological phase transition, which we believe is due to the finite size effect, as discussed above.

Finally, we note that in a more general parameter regime with $r\neq\sin{\theta}$, the non-Bloch Hamiltonian becomes more complicated as now the SSH analog no longer holds, and an analytical solution of the topological phase transition is difficult to obtain even omitting the effect of $h_0$. Nevertheless, compared with the previous cases, numerically we do not find any qualitatively different topological property in this general scenario.

\section{Summary}\label{V}
In summary, we have comprehensively investigated the topological properties of a non-Hermitian Creutz ladder model with non-Hermiticity added in several different manners, including asymmetric amplitudes of each hopping parameter and on-site imaginary potentials. 
Besides carefully analyzing the topological phase transitions between gapped and gapless phases in each scenario,
we further unveil that various phenomena can emerge depending on how the non-Hermiticity enters the Hamiltonian,
including a winding topology in a 2D plane associated with a complex Hamiltonian vector, the emergence of a high spectral winding number without introducing long-range hoppings, and a competition between two types of NHSE. 
Additionally, we find that the NHSE does not necessarily lead to the breakdown the conventional BBC, which is also associated with whether EPs emerge in the concerned parameter regime.

\section{Acknowledgement}
The work is supported
by the Guangdong Basic and Applied Basic Research
Foundation (No. 2020A1515110773).
\appendix

\section{Transfer matrix approach for the case with $\alpha$}\label{app:TM_alpha}
As discussed in the main text, the system with a nonzero $\alpha$ is free of the NHSE, thus the topological edge modes under the OBC shall be directly associated with the a phase winding of eigenmodes of the Bloch Hamiltonian $h(k)$ in Eq. \eqref{eq:H_alpha}. Here we first omit the $h_0(k)$ term as it does not affect the eigenmodes of $h(k)$. 
The possible topological edge modes are now fixed at zero-energy, as the resulting Hamiltonian $h'(k)=h(k)-h_0(k)$ satisfies a sublattice symmetry $$\sigma_yh'(k)\sigma_y=-h'(k).$$
We further apply a rotation of the Pauli's matrices, $\sigma_z\rightarrow\sigma_y\rightarrow-\sigma_z$, 
which decouples the two components in the transfer matrix approach for the zero-energy eigenmodes, as discussed later.
The final Bloch Hamiltonian becomes
\begin{eqnarray}
h'_{\rm rot}(k)=(M+2r\cos k)\sigma_x-\delta r\sin k\sigma_y,
\end{eqnarray}
with $\delta r=\left(e^\alpha+e^{-\alpha}\right)\sin{\theta}-i\left(e^{-\alpha}-e^{\alpha}\right)\cos{\theta}$.

In the real-space, the Hamiltonian is rewritten as
\begin{eqnarray}
\hat{H}' _{\rm rot}&=& \sum_n\left[M\hat{a}^\dag_n\hat{b}_n+ \left(r-\frac{1}{2}\delta r\right)\hat{a}^\dag_n\hat{b}_{n+1}\right. \nonumber\\
&&\left.+\left(r+\frac{1}{2}\delta r\right)\hat{b}^\dag_{n}\hat{a}_{n+1}\right]+\text{h.c.}
\end{eqnarray}
By solving the Schr\"odinger equation $\hat{H}' _{\rm rot}\Psi=E\Psi$ with $\Psi=\sum_{x}[\psi_{x}^A\hat{a}_x^\dagger+\psi_{x}^B\hat{b}_x^\dagger]$ an eigenmode of the system, we obtain the recursive conditions
\begin{eqnarray}
	M\psi_{x}^A+R_-\psi_{x-1}^A+R_+\psi_{x+1}^A &=& E_n\psi_{x}^B\\
	M\psi_{x}^B+R_-\psi_{x+1}^B+R_+\psi_{x-1}^B &=& E_n\psi_{x}^A
\end{eqnarray} 
where $R_{\pm} = r\pm\delta r/2$. As the concerned topological edgemodes are fixed at the zero-energy due to the Chiral symmetry $h'(k)$, we set $E=0$, thus the eigenmode amplitudes $\psi_{x}^{A,B}$ for the two components are decoupled from either other.
Hence we can write the recursive conditions in the form of the transfer matrix \cite{sanchez2012transfer}, given by
\begin{eqnarray}
	\begin{pmatrix}
		\psi_{x+1}^A	\\
		\psi_x^A
	\end{pmatrix}
	&=&T_A\begin{pmatrix}
		\psi_x^A\\
		\psi_{x-1}^A
	\end{pmatrix}
	=
	\begin{pmatrix}
	-\frac{M}{R_+} 	&-\frac{R_-}{R_+}\\
	1	&0
	\end{pmatrix}
	\begin{pmatrix}
		\psi_x^A\\
		\psi_{x-1}^A
	\end{pmatrix},
	\nonumber\\\label{eq:TA}
	\\
		\begin{pmatrix}
		\psi_{x+1}^B	\\
		\psi_x^B
	\end{pmatrix}
	&=&T_B\begin{pmatrix}
		\psi_x^B\\
		\psi_{x-1}^B
	\end{pmatrix}
	=
	\begin{pmatrix}
	-\frac{M}{R_-} 	&-\frac{R_+}{R_-}\\
	1	&0
	\end{pmatrix}
	\begin{pmatrix}
		\psi_x^B\\
		\psi_{x-1}^B
	\end{pmatrix}.\nonumber\\\label{eq:TB}
\end{eqnarray}

\begin{figure}
\includegraphics[width=1\linewidth]{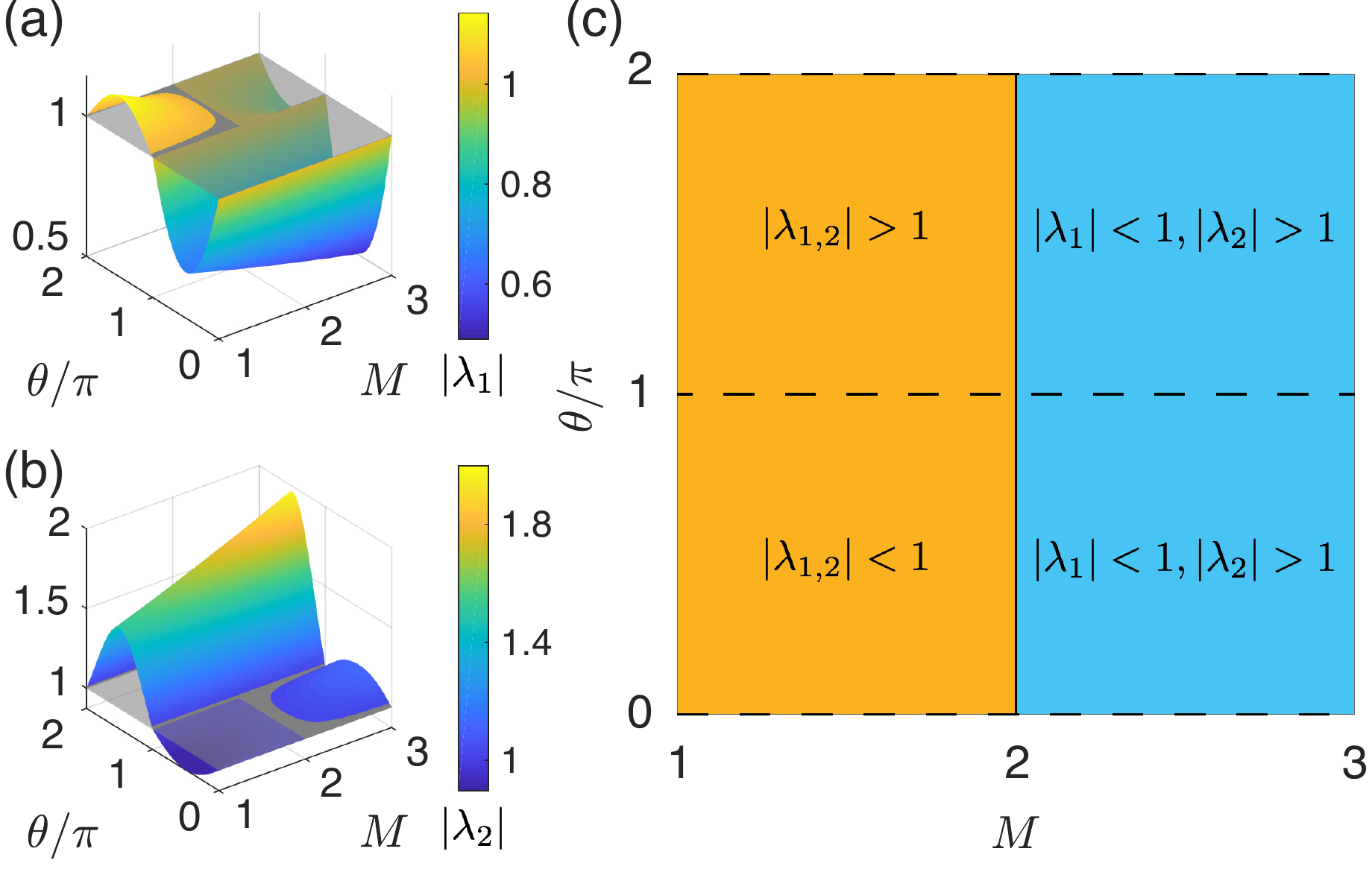}
\caption{(a) and (b) the absolute values of $\lambda_{1,2}$, the two eigenvalues of the transfer matrix $T_A$ respectively. The gray translucent plane represents $|\lambda_{1,2}|=1$. (c) A phase diagram of the system, determined by the numerical results of $|\lambda_{1,2}|$. The yellow region with $|\lambda_{1,2}|<1$  ($|\lambda_{1,2}|>1$) means that an zero-energy eigen-solution localizing at the $x=1$ ($x=L$) exists for the $A$ component, and another one localizing at $x=L$ ($x=1$) exists for the $B$ component. 
The blue region with $|\lambda_1|<1$ and $|\lambda_2|>2$ corresponds to no zero-energy eigen-solution.
Note that due to the rotation of the pseudospin, here $A$ and $B$ are different combinations of the two sublattices of the Creutz ladder.
Specially, $|\lambda_{1,2}|=1$ along the dash lines at $\theta=0~(2\pi)$ and $\pi$.
Other parameters are $\alpha=2$ and $r=1$.
}
\label{fig:alpha_TM}
\end{figure}

To find out the condition for the system to support zero-energy edgemodes,
we take $T_A$ as an example and consider a semi-infinite boundary condition, i.e. with $x\in[1,\infty)$.
The vector $(\psi_1^A, 0)^T$ (as $\psi_0^A=0$ is determined by the boundary condition) can be rewritten as a linear superposition of the eigenvectors of matrix $T_A$, 
$$(\psi_1^A, 0)^T = c(a_1,b_1)^T + s(a_2,b_2)^T,$$
with $T_A(a_i,b_i)^T=\lambda_i(a_i,b_i)^T$.
Substituting it to Eq.\eqref{eq:TA}, we obtain
\begin{equation}
\begin{aligned}
	(\psi_{x+1}^A, \psi_{x}^A)^T = c\lambda_1^L(a_1,b_1)^T + s\lambda_2^L(a_2,b_2)^T,\\
\end{aligned}
\end{equation}
meaning that $\psi_{x}^A$ decays exponentially and is vanishing with $x\rightarrow\infty$ when $|\lambda_{1,2}|<1$. 
That is,  we can now normalize this eigen-solution by requiring $\sum^L_{x=1} |\psi_x^A|^2=1$, and obtain an edgemode localized at $x=1$. Similarly, another zero-energy edgemode localized at $x=L$ can be obtained from the transfer matrix $T_B$ for  component B, by considering the other semi-infinite boundary condition $x\in(-\infty,L]$. The same conclusion can also be obtained when $|\lambda_{1,2}|<1$, where $T_A$ leads to a zero-energy edgemode localized at $x=L$, and $T_B$ leads to the other one localized at $x=1$.
However, other then these two conditions, the eigen-solution cannot be normalized, corresponding to the absence of zero-energy edgemode \cite{li2015hidden,sanchez2012transfer}.

In our model, 
the eigenvalues of the transfer matrix $T_A$ are given by
\begin{equation}
	\lambda_{1,2} = \frac{-M\pm\sqrt{M^2-4R_+R_-}}{2R_+}.\label{eq:lambda12}
\end{equation}
A simple solution can be obtained when $M=2r$, where 
\begin{eqnarray}
\lambda_1=\frac{-2r+\delta r}{2r+\delta r},\lambda_2=-1.
\end{eqnarray}
Here we have chosen $\sqrt{\delta r^2}=\delta r$ for simplicity, as the two eigenvalues correspond to $\pm\sqrt{\delta r^2}$ respectively.
Apparently, $|\lambda_1|\neq 1$ is generally not satisfied, therefore $\lambda_2=-1$ suggests that $M=2r$ is a topological transition point of the model. This is further confirmed by numerically solving $\lambda_{1,2}$ from Eq. \eqref{eq:lambda12}, as shown in Fig. \ref{fig:alpha_TM}.
It is seen that the eigenvalues satisfy $|\lambda_{1,2}|<1$ or $|\lambda_{1,2}|>1$ only when $|M|<|2r|$, in consistence with the results and discussion in the main text.

\section{Deriving the phase boundaries of the case with nonzero $m$ and $r_1$}\label{app:m_r1}
Following the discussion in the main text, the model with non-Hermiticity given by $m$ and $r_1$ exhibits no NHSE, 
and its phase transitions can be determined solely from the PBC Hamiltonian matrix $h(k)$.
Specifically, the band-touching point is given by $P(k)=0$, with
\begin{eqnarray}
P(k)&=&h_x^2+h_y^2+h_z^2\nonumber\\
&=&(M+2r\cos{k})^2+(-2\sin{\theta}\sin{k})^2\nonumber\\
&&-(m+2r_1\cos{k})^2.\nonumber
\end{eqnarray}
As $P(k)$ always takes real values, we may only look at the extreme points of $P(k)$, which are
\begin{equation}
\begin{aligned}
	P(0) &= M^2-m^2+4(r^2-r_1^2)+4(Mr-mr_1)\\
	P(\pi) &= M^2-m^2+4(r^2-r_1^2)-4(Mr-mr_1)
\end{aligned}
\end{equation}
and
\begin{equation}
	P(k') = M^2-m^2+4\sin^2{\theta}+\frac{(Mr-mr_1)^2}{\sin^2{\theta}-r^2+r_1^2},\label{eq:app_Pkprime}
\end{equation}
with
\begin{equation}
	\cos k' = \frac{Mr-mr_1}{2(\sin^2{\theta}-r^2+r_1^2)}.
\end{equation}
Note that the third extreme point exists only when 
\begin{eqnarray}
|\frac{Mr-mr_1}{2(\sin^2{\theta}-r^2+r_1^2)}| \leqslant1.\label{eq:app_coskprime}
\end{eqnarray}
Therefore, the system has a real energy spectrum with a real line-gap when
\begin{eqnarray}
P(0),~P(\pi),~P(k')>0,
\end{eqnarray}
an imaginary one with a imaginary line-gap when
\begin{eqnarray}
P(0),~P(\pi),~P(k')<0,
\end{eqnarray}
which are the same for this model as the conditions of Eqs. \eqref{eq:condition1_m_r1} and \eqref{eq:condition2_m_r1} in the main text.
Otherwise, $P(k)=0$ must be satisfied for certain values of $k$, as the $P(k)$ take different signs at different extreme points.
In such a circumstance, the eigenenergies distribute along both the imaginary and real axises in the complex plane, and form a gapless cross-shape spectrum [see Fig. \ref{fig:phase_m_r1}(c2) and (d2) for some examples].

\begin{figure}
\includegraphics[width = 1\linewidth]{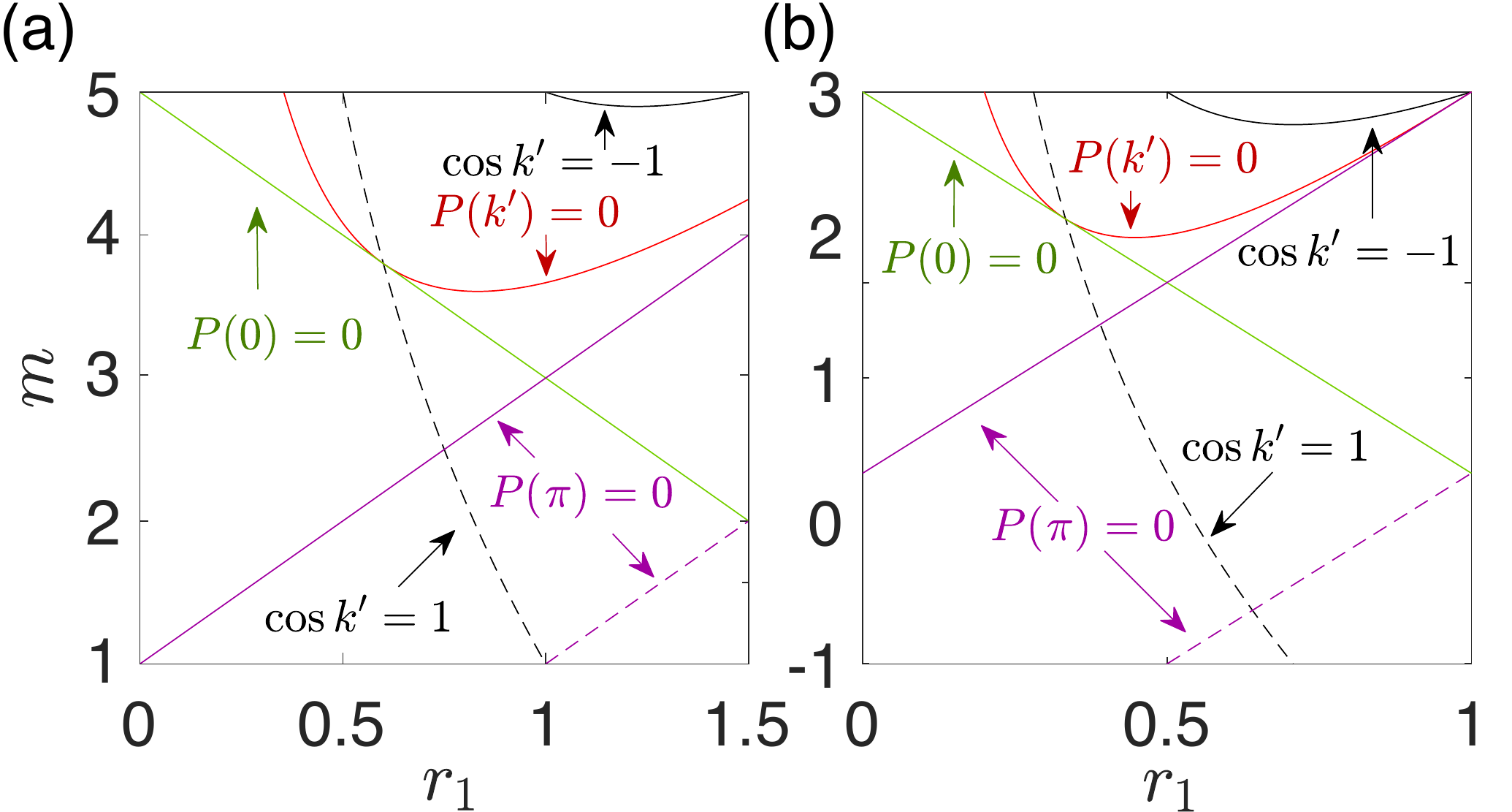}
\caption{Results from the conditions of Eqs. \eqref{eq:app_boundaries1_m_r1} and \eqref{eq:app_boundaries2_m_r1}, with (a) $M=3$, $r=1$, $\theta=\pi/2$ and (b) $M=r=1$, $\theta=\pi/2$, the same as the parameters for Panels (a) and (b) in Fig. \ref{fig:phase_m_r1} respectively.
}
\label{fig:boundaries_m_r1}
\end{figure}

The phase diagrams in Fig. \ref{fig:phase_m_r1} are obtained from the above conditions with the given parameters. That is, for the parameters $M=3$, $r=1$, $\sin{\theta}=1$ as that of Fig. \ref{fig:phase_m_r1}(a), the possible phase boundaries are given by 
\begin{eqnarray}
P(0) = 0& \quad\Rightarrow\quad & m_\pm = \pm 5-2r_1,\\
P(\pi) = 0& \quad\Rightarrow\quad & m_\pm = \pm 1+2r_1,\\
P(k')= 0& \quad\Rightarrow\quad & m = \frac{13r_1^2+9}{6r_1},\\
\cos k'=\pm1& \quad\Rightarrow\quad & m_\pm = \frac{3\mp 2r_1^2}{r_1}.\label{eq:app_boundaries1_m_r1}
\end{eqnarray}
The parameters $M=r=1$, $\sin{\theta}=1$ for Fig. \ref{fig:phase_m_r1}(b) lead to
\begin{eqnarray}
P(0) = 0& \quad\Rightarrow\quad & m_\pm = \pm 3-2r_1,\\
P(\pm\pi) = 0& \quad\Rightarrow\quad & m_\pm = \pm 1+2r_1,\\
P(k')= 0& \quad\Rightarrow\quad & m = \frac{5r_1^2+1}{2r_1},\\
\cos{k'} = \pm 1 & \quad\Rightarrow\quad & m_\pm = \frac{1\pm 2r_1^2}{r_1}.\label{eq:app_boundaries2_m_r1}
\end{eqnarray}
The lines and curves corresponding to these conditions are shown in Fig.\ref{fig:boundaries_m_r1}(a) and (b) respectively, and the phase diagrams are obtained by further checking whether $P(k)$ takes positive or negative values within different areas separated by these lines and curves.

For a more general parameter regime with $\theta\neq\pi/2$, $h_0$ becomes nonzero and may lead to another gapless phase with the two bands overlap at different $k$ values.
Here we shall consider the phase transition induced by $\theta$ for the two specific cases in Fig. \ref{fig:spectrum_m_r1_thetaNeq} as illustrations. The first one of Fig. \ref{fig:spectrum_m_r1_thetaNeq}(a) and (c) is with the parameters
\begin{eqnarray}
M=3,r=1,m=2.5,r_1=1,
\end{eqnarray}
resulting in 
$$P(k)=5+4\cos{k}+4\sin^2{k}\sin^2{\theta}>0$$
for arbitrary $k$, meaning that the spectrum is always real.
The overlapping condition of the two bands is given by $$E_-(k)_{\rm max}>E_+(k)_{\rm min}$$, 
which reads
\begin{equation}
	E_-(0)>E_+(\pi)\quad\Rightarrow\quad\cos{\theta}>\frac{\sqrt{0.75}+\sqrt{4.75}}{4},
\end{equation}
for the given parameters, after some careful analysis of the extreme values of $E_\pm(k)=h_0\pm\sqrt{P(k)}$.
The phase transition point predicted by $E_-(0)=E_+(\pi)$, i.e. $\theta\approx\pm0.2245\pi$, is in good consistent with the numerical results in Fig. \ref{fig:spectrum_m_r1_thetaNeq}(a) and (b).

Similarly, the other case in Fig. \ref{fig:spectrum_m_r1_thetaNeq}(c) and (d) with
\begin{eqnarray}
M=1,r=1,m=1.5,r_1=0.5
\end{eqnarray}
has
$$P(k)=-1.25+3\cos^2{k}+\cos{k}+4\sin^2{k} \sin^2{\theta},$$
and the overlapping condition reads
\begin{equation}
	E_-(0)>E_+(\pi)\quad\Rightarrow\quad \cos{\theta}>\frac{\sqrt{0.75}+\sqrt{2.75}}{4},
\end{equation}
predicting a phase transition at $\theta\approx\pm0.2826\pi$ [see Fig. \ref{fig:spectrum_m_r1_thetaNeq}(c) in the main text].
Furthermore, now that $P(k)$ takes negative values for certain values of $k$ and $\theta$, the system also supports a band-touching gapless phase in this parameter regime. The band-touching condition requires $P(k')<0$, as $P(0)=2.75>0$, $P(\pi)=0.75>0$ at the other two extreme points of $P(k)$. Therefore we obtain
\begin{equation}
	-1.25+4\sin^2{\theta}+\frac{0.0625}{\sin^2{\theta}-0.75}<0\\
\end{equation}
\begin{equation}
	|\frac{0.25}{2\sin^2{\theta}-1.5}|<1
\end{equation}
from Eqs. \eqref{eq:app_Pkprime} and \eqref{eq:app_coskprime} respectively.
Combining these two conditions,
we reach the conclusion that the two band touches and some eigenenergies acquire nonzero imaginary values when $|\sin{\theta}|<0.59307$, resulting in a transition point at $\theta\approx0.2021\pi$ [see Fig. \ref{fig:spectrum_m_r1_thetaNeq}(c) and (d) in the main text].

\section{Conditions for preserving the conventional BBC in the presence of the NHSE}\label{app:BBC_NHSE}
In this section we shall discuss the conditions for preserving the conventional BBC even in the presence of the NHSE.
Specially, in order to violate the conventional BBC, the system must support exceptional points (EPs) between different bands at certain parameters. To see this, we first review the conditions for having the NHSE, and how the conventional BBC may be violated in non-Hermitian systems.

\begin{figure}
\includegraphics[width=1\linewidth]{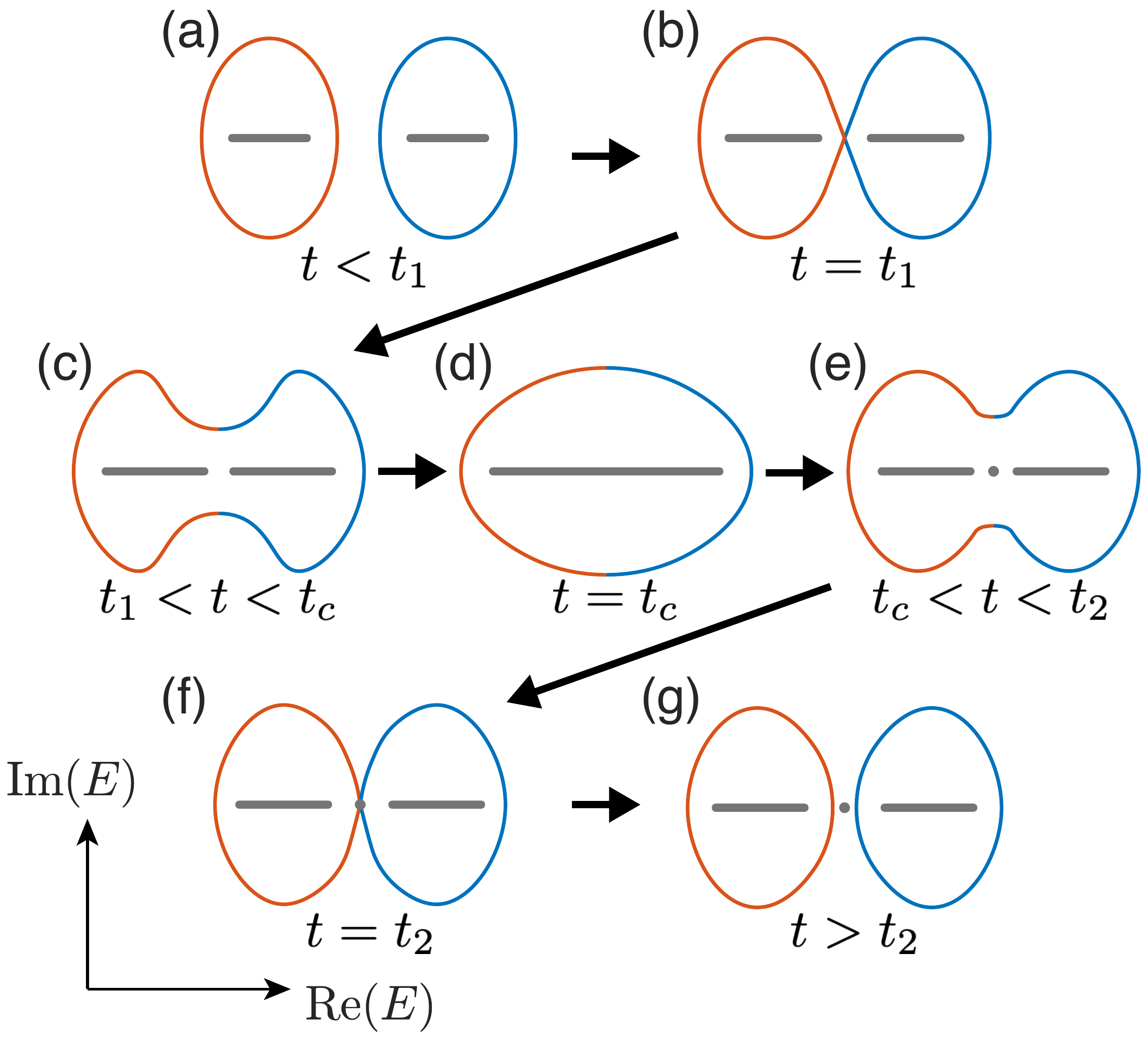}
\caption{Sketches for topological transition of the PBC (blue and red for the two bands) and OBC (gray) spectra when the conventional BBC is broken. 
$t$ represents a parameter which drives the system to undergo a topological transition.
The two PBC bands are assumed to touch each other at (b) $t=t_1$ and (f) $t=t_2$, and connect into a big loop (but without any degeneracy at any individual crystal momentum) for $t_1<t<t_2$.
The OBC gap may close only at $t_c\in(t_1,t_2)$, as the OBC spectrum must fall within the interior of the PBC one.
The conventional BBC is broken as now the PBC gap-closing point is different from the OBC one.
}
\label{fig:sketch_topo}
\end{figure}

\begin{figure}
\caption{Sketches for topological transition of the PBC (blue and red for the two bands) and OBC (gray) spectra when the conventional BBC is kept. 
$t$ represents a parameter which drives the system to undergo a topological transition.
The two PBC bands are assumed to touch each other at (b) $t=t_c$, and remains two separated bands otherwise. The OBC gap closing point must coincide with that under PBC, as the OBC spectrum must fall within the interior of the PBC one.
}
\label{fig:sketch_topo2}
\end{figure}

(i) For a non-Hermitian system, the NHSE has a correspondence to a loop-like PBC spectrum enclosing a nonzero area in the complex plane, and its OBC spectrum must lie within this area enclosed by the PBC spectrum \cite{Lee2019anatomy,lee2020unraveling,zhang2019correspondence,okuma2020topological}, as shown by the sketches in Fig. \ref{fig:sketch_topo}.

(ii) In a two-band non-Hermitian system with the NHSE, the two bands may connect into a big loop in certain parameter regimes, say for a parameter $t$ ranging from $t_1$ to $t_2$, as shown by Fig. \ref{fig:sketch_topo}(c)-(e). 

(iii) Its OBC spectrum is generally still given by two separated bands, yet the gap between them may close at certain $t_c\in(t_1,t_2)$, corresponding to the topological phase transition under OBC (Fig. \ref{fig:sketch_topo}(d)). However, the PBC spectrum has a band touching point only at $t=t_1$ or $t=t_2$, and shows no transition at $t=t_c$. Therefore the OBC topological phase transition at $t_c$ cannot be predicted by the Bloch Hamiltonian of the PBC system, which is known as the breakdown of the conventional BBC.

Because of (i), a key condition for violating the conventional BBC is that the spectrum supports only point-gaps, but no a line-gap between the two concerned energy bands. In the above scenario, this condition is to have the two bands connecting into a single loop, where an adiabatic following of the eigenstates goes back to the initial state only after the momentum $k$ varies two periods, e.g. from $0$ to $4\pi$ \cite{xiong2018does}.
Such a scenario will give an overall Berry phase of $\gamma=\pi$ for the two bands, meaning that the trajectory of the Hamiltonian vector must cycle an EP \cite{Mailybaev2005geometric,shen2018topological,Yin2018nonHermitian,li2019geometric}.
On the other hand, no EP is enclosed by this trajectory when the two PBC bands are separated, as the summed Berry phase of them must be $\gamma=0$.  This is also always the case for Hermitian gapped systems.
Therefore, when turning on certain non-Hermitian parameters, the conventional BBC may be violated only when the system goes into the former regime with $\gamma=\pi$, meaning that the system must goes through an exceptional gap-closing at some point, so that the trajectory of the Hamiltonian vector can enclose an EP.

Finally, we note that it is also possible to have a spectrum with only point-gaps by having the two energy bands overlapping with each other at different values of $k$. This condition requires no any type of degeneracies between the two bands. 
However, we do not find such an example with the conventional BBC being violated.

\end{document}